\documentclass[conference]{IEEEtran}
\usepackage[hidelinks]{hyperref}

\usepackage{caption}
\usepackage{subcaption}

\usepackage[framemethod=tikz]{mdframed}
\usepackage[most]{tcolorbox}
\usepackage[normalem]{ulem}
\usepackage{algorithm}
\usepackage{algpseudocode}
\usepackage{amsfonts}
\usepackage{amsmath}
\usepackage{arydshln}
\usepackage{balance}
\usepackage{blindtext}
\usepackage{booktabs}
\usepackage{cancel}
\usepackage[font=footnotesize]{caption}
\usepackage{color,soul}
\usepackage{comment}
\usepackage{enumitem}
\usepackage{fontawesome}
\usepackage{graphicx}
\usepackage{xcolor}
\usepackage{listings}
\usepackage{mathtools}
\usepackage{multirow}
\usepackage{pifont}
\usepackage{tabularx}
\usepackage{threeparttable}
\usepackage{tikz}
\usepackage{colortbl}
\usepackage{xspace}
\usepackage{wasysym}

\usepackage{makecell}
\usepackage{svg}

\definecolor{winered}{rgb}{0.7,0,0}
\definecolor{gray}{gray}{0.7}
\definecolor{darkpastelgreen}{rgb}{0.01, 0.75, 0.24}
\definecolor{cadmiumgreen}{rgb}{0.0, 0.42, 0.24}
\definecolor{brickred}{rgb}{0.8, 0.25, 0.33}
\definecolor{cornellred}{rgb}{0.7, 0.11, 0.11}
\definecolor{burgundy}{rgb}{0.5, 0.0, 0.13}
\definecolor{frenchblue}{rgb}{0.0, 0.45, 0.73}
\definecolor{light-gray}{gray}{0.92}
\definecolor{lightlight-gray}{gray}{0.97}
\definecolor{codegray}{gray}{0.90}
\definecolor{inputgray}{gray}{0.90}
\definecolor{darkgreen}{RGB}{40,125,40}

\hypersetup
{
colorlinks=true,
linkcolor=winered,
urlcolor={winered},
filecolor={winered},
citecolor={winered},
allcolors={winered}
}

\newcommand{\cmt}[1]{}

\newcommand{\homedir}{\raise.17ex\hbox{$\scriptstyle\sim$}}

\newcommand{\etal}{et al.\xspace}

\global\mdfdefinestyle{rtboxstyle}{%
linecolor=black,%
leftmargin=0cm,rightmargin=0cm,linewidth=0.5pt,
roundcorner=3,
skipbelow=0pt,backgroundcolor=lightlight-gray
}

\newcommand{\code}[1]{\texttt{#1}}

\mdfdefinestyle{MyFrame}{%
     roundcorner=1pt,
     innertopmargin=6pt,
     innerbottommargin=6pt,
     innerrightmargin=6pt,
     innerleftmargin=6pt,
    }

\graphicspath{{figure/}{figures/}}
\DeclareGraphicsExtensions{.pdf,.eps,.png,.jpg,.jpeg}

\long\def\comment#1{}

\renewcommand{\paragraph}[1]{\smallskip\noindent\emph{#1}\quad}
\def\Snospace~{\S{}}

\newcommand{\heading}[1]{{\vspace{2pt}\noindent\bf{#1}}} 
\newcommand{\headinggi}[1]{{\em{#1}}} 

\newcommand{\eg}{{\it e.g.,~}}
\newcommand{\ie}{{\it i.e.,~}}
\newcommand{\etc}{{\it etc.~}}

\newcommand{\nm}[1]{{\it (#1)\xspace}} 

\newcommand{\optional}[1]{}

\def\pname{{\textsc{ProvExplainer}}\xspace}
\usepackage{fp}

\newif\ifbiblatex
\biblatextrue

\newif\ifcomments
\commentstrue 

\ifcomments
\usepackage{todonotes}
	\newcommand{\kjee}[1]{\textcolor{red}{[KJEE: #1]}}
	\newcommand{\murat}[1]{\textcolor{blue}{[MK: #1]}}
    \newcommand{\mh}[1]{\textcolor{blue}{[MH: #1]}}
	\newcommand{\wc}[1]{}
	\newcommand{\kunal}[1]{\textcolor{orange}{[{\bf KUNAL:} #1]}}
	\newcommand{\jdw}[1]{\textcolor{orange}{[{\bf JDW:} #1]}}
	\newcommand{\fix}[1]{\textcolor{red}{\hl{#1}}}
	\newcommand{\corr}[2]{\sout{#1} \hl{#2}}
\else
	\newcommand{\kjee}[1]{}
	\newcommand{\murat}[1]{}
	\newcommand{\mh}[1]{}
	\newcommand{\wc}[1]{}
	\newcommand{\kunal}[1]{}
	\newcommand{\jdw}[1]{}
	\newcommand{\fix}[1]{}
	\newcommand{\corr}[2]{}
\fi

\def\gnne{GNNExplainer\xspace}%
\def\subx{SubgraphX\xspace}%
\def\pge{PGExplainer\xspace}%

\def\trusteedt{Surrogate DT\xspace}%

\def\sage{GraphSAGE~\xspace}
\def\gat{GAT~\xspace}

\def\fm{Fileless Malware~\xspace}
\def\fidelityp{$\text{fidelity}^{+}$~\xspace}
\def\fidelitym{$\text{fidelity}^{-}$~\xspace}
\def\darpatc{DARPA TC~\xspace}

\RequirePackage[normalem]{ulem} 
\RequirePackage{color}\definecolor{RED}{rgb}{1,0,0}\definecolor{BLUE}{rgb}{0,0,1} 
\providecommand{\DIFdel}[1]{{\protect\color{red}\sout{#1}}}                      
\providecommand{\DIFdel}[1]{}                      
\usepackage[acronym]{glossaries}

\newacronym{hdl}{HDL}{High-level Dynamic Language}

\newacronym{ml}{ML}{Machine Learning}
\newcommand{\ml}{\gls*{ml}\xspace}

\newacronym{ai}{AI}{Artificial Intelligence}

\newacronym{nn}{NN}{Neural Network}

\newcommand{\nns}{\glspl*{nn}\xspace}

\newacronym{gnn}{GNN}{Graph Neural Network}
\newcommand{\gnn}{\gls*{gnn}\xspace}
\newcommand{\gnns}{\glspl*{gnn}\xspace}

\newacronym{dnn}{DNN}{Deep Neural Network}

\newacronym{rnn}{RNN}{Recurrent Neural Network}

\newacronym{llm}{LLM}{Large Language Model}

\newacronym{mlm}{MLM}{Masked Language Model}

\newacronym{gan}{GAN}{Generative Adversarial Network}

\newacronym{nlp}{NLP}{Natural Language Processing}

\newacronym{pl}{PL}{Programming Language}

\newacronym{vm}{VM}{Virtual Machine}

\newacronym{bert}{BERT}{Bidirectional Encoder Representations from Transformers}

\newacronym{csn}{CSN}{Code Search Net}

\newacronym{sota}{SOTA}{State-Of-The-Art}
\newcommand{\sota}{\gls*{sota}\xspace}

\newacronym{pypi}{PyPI}{Python Package Index}

\newacronym{cdm}{CDM}{Common Data Model}
\newcommand{\cdm}{\gls*{cdm}\xspace}

\newacronym{tc}{TC}{Transparent Computing}

\newacronym{dt}{DT}{Decision Tree}
\newcommand{\dt}{\gls*{dt}\xspace}
\newcommand{\dts}{\glspl*{dt}\xspace}

\newacronym{apt}{APT}{Advanced Persistent Threat}
\newcommand{\apt}{\gls*{apt}\xspace}
\newcommand{\apts}{\glspl*{apt}\xspace}

\newacronym{cve}{CVE}{Common Vulnerabilities and Exposures}

\newacronym{ids}{IDS}{Intrusion Detection System}
\newcommand{\ids}{\gls*{ids}\xspace}
\newcommand{\idss}{\glspl*{ids}\xspace}

\newacronym{av}{AV}{Anti-Virus}

\newacronym{edr}{EDR}{Endpoint Detection and Response}
\newcommand{\edr}{\gls*{edr}\xspace}

\newacronym{poi}{POI}{Point-Of-Interest}

\newacronym{vlan}{VLAN}{Virtual Local Area Network}

\newacronym{ctwo}{C2}{Command and Control}

\newacronym{csg}{CSG}{Computer Security Group}

\newacronym{wicys}{WiCyS}{Women In Cyber Security}

\newacronym{nsa}{NSA}{National Security Agency}

\newacronym{dod}{DoD}{Department of Defense}

\newacronym{lsm}{LSM}{Log Structured Merge}

\newacronym{me}{ME}{Microelectronics}

\newacronym{ci}{CI}{CyberInfrastructure}

\newacronym{soc}{SoC}{System-on-Chip}

\newacronym{cots}{COTS}{Custom-Off-The-Shelves}

\newacronym{tle}{TLE}{Two-line element set}

\newacronym{wal}{WAL}{Write Ahead Log}

\newacronym{obc}{OBC}{On-Board Computer}

\newacronym{dos}{DoS}{Denial-of-Services}

\newacronym{ttp}{TTP}{Tactics, techniques, and Procedures}
\newcommand{\ttp}{\gls*{ttp}\xspace}
\newcommand{\ttps}{\glspl*{ttp}\xspace}

\newacronym{cti}{CTI}{Cyber Threat Intelligence}
\newcommand{\cti}{\gls*{cti}\xspace}

\newacronym{ioc}{IOC}{Indicators Of Compromise}

\newacronym{jspoc}{JSpOC}{The Joint Space Operations Center}

\newacronym{cfg}{CFG}{Control Flow Graph}

\newacronym{cdg}{CDG}{Control Dependency Graph}

\usepackage[utf8]{inputenc}
\usepackage[english]{babel}
\usepackage{csquotes}
\usepackage{multicol}
\usepackage{titling}
\usepackage{balance}
\usepackage{listings}
\usepackage{tikz}
\usepackage{amsmath}
\usepackage{times,url,color,soul,xspace,enumitem}
\usepackage{graphicx}
\usepackage{caption}
\usepackage{comment}
\usepackage[inline,draft,nomargin,index]{fixme}

\usepackage[
  backend=biber,
  style=ieee,
  citestyle=numeric-comp,
  sorting=none,   
  sortcites=true, 
  giveninits=true,
  mincrossrefs=99
]{biblatex}

\usepackage{dsfont}
\usepackage{multicol}
\graphicspath{{figs/}}

\begin{document}

\date{}


\def\papertitle{Interpreting GNN-based IDS Detections Using Provenance Graph Structural Features}
\title{\papertitle}

\IEEEoverridecommandlockouts
\makeatletter\def\@IEEEpubidpullup{6.5\baselineskip}\makeatother
\IEEEpubid{\parbox{\columnwidth}{
    Network and Distributed System Security (NDSS) Symposium 2024\\
    26 February - 1 March 2024, San Diego, CA, USA\\
    ISBN 1-891562-93-2\\
    https://dx.doi.org/10.14722/ndss.2024.23xxx\\
    www.ndss-symposium.org
}
\hspace{\columnsep}\makebox[\columnwidth]{}}

\author{
{\rm Kunal Mukherjee$^{\dagger}$,}
{\rm Joshua Wiedemeier$^{\ddagger}$,}
{\rm Tianhao Wang$^{\ddagger}$,}
{\rm Feng Chen$^{\ddagger}$,}
{\rm Muhyun Kim$^{\ddagger}$,}
\\
{\rm Murat Kantarcioglu$^{\dagger}$, and}
{\rm Kangkook Jee$^{\ddagger}$} \\
\\
$^{\dagger}${\rm Virginia Tech} 
$^{\ddagger}${\rm The University of Texas at Dallas}
} 




\maketitle
\begin{abstract}\label{sec:abstract}
    Advanced cyber threats (\eg \fm and \apt) have driven the adoption of \textit{provenance}-based security solutions. These solutions employ ML models for behavioral modeling and critical security tasks such as malware and anomaly detection. However, the opacity of ML-based security models limits their broader adoption, as the lack of transparency in their decision-making processes restricts explainability and verifiability. We tailored our solution towards \gnn-based security solutions since recent studies employ GNNs to comprehensively digest system provenance graphs for security critical tasks. 
    
    To enhance the \textit{explainability} of GNN-based security models, we introduce \pname, a framework offering \textit{instance-level} \textit{security-aware} explanations using an interpretable surrogate model. \pname's interpretable feature space consists of \textit{discriminant subgraph patterns} and \textit{graph structural features} which can be directly mapped to the system provenance problem space, making the explanations human understandable. Considering both the subgraph patterns and graph structural features, gives \pname the unique advantage of providing explanations that are sensitive to both local and global contexts. 
    
    By considering prominent \gnn architectures (\eg \gat and \sage) for anomaly detection tasks, we show how \pname outperformed the current \sota \gnn explainers to deliver security domain and instance-specific explanations. We measure the explanation quality using \fidelityp/ \fidelitym metric as used by traditional \gnn explanation literature, we incorporate the precision/recall metric where we consider the accuracy of the explanation against the ground truth, and we designed a human actionability metric based on graph traversal distance. On real-world \fm and APT datasets, \pname achieves up to 29\%/27\%/25\%/1.4x higher \fidelityp, precision, recall and actionability (where higher values are better), and 12\% lower \fidelitym (where lower values are better) when compared against \sota \gnn explainers. 

\end{abstract}
\section{Introduction}\label{sec:introduction}

Given the significant threat posed by advanced and sophisticated adversaries~\cite{apt0,apt1,sony,solarwinds} (\ie malware writers and \apt actors), security is paramount in today's highly digitized society. Recent advances in system monitoring have generated a variety of fine-grained telemetry for advanced security analysis. To leverage this telemetry data to counter threats, various learning-based security tools~\cite{dahl2013large, arp2014drebin, gandotra2014malware, saxe2015deep, grosse2016adversarial, narayanan2016performance, kolosnjaji2016deep, javaid2016deep, tang2016deep,holmes2019sp, wang2020ndss, han2020ndss, morse2020sp, sigl2021sec, cheng2024kairos, rehman2024flash, goyal2024rcaid} have been deployed, with provenance-based solutions being the most notable. 

System provenance captures causally dependent system events on an host, represented as a provenance graph. Provenance graphs are heterogeneous, comprising various node types (\eg process, file, and socket) and edge types (\eg read, write, and send), each with textual and numerical attributes, effectively representing dynamic runtime behaviors. Hence, provenance graphs provide irreplaceable security defense for countering sophisticated and stealthy APT attack campaigns. Particularly, \edr solutions~\cite{xdr0,xdr1,edr0}, built on fine-grained system provenance data, have become a mainstream security defense for enterprises.

Recent advancements in \gnns enable the direct learning of relational dependencies and topological structures from graphs.
\gnns utilize a permutation-invariant message-passing mechanism to aggregate information from node neighborhoods, expanding the information propagated through each neighborhood with every iteration. This allows \gnns to learn from complex graph structures automatically, which has proven beneficial for provenance solutions~\cite{goyal2024rcaid, rehman2024flash, cheng2024kairos, jia2024magic, bilot12399sometimes, jian2025}.

While \gnn-based security models are highly effective, their opaque nature significantly limits their advantages, \ie due to the lack of clear and verifiable explanations. These models fail to provide \textit{ground truth relevant} explanations, which are essential for building trust among security practitioners. In this context, ground truth relevant explanations refer to explanations aligned with specific system actions that represent \ttps, and system artifacts that are familiar to analyst and can be validated by existing security intelligence (\eg attack memos~\cite{darpa:ground, darpa:ground2} and VirusTotal~\cite{virustotal} reports).

While studies~\cite{ying2019gnnexplainer, luo2020parameterized, yuan2021explainability} have explored domain-agnostic explanations for \gnn decisions, they lack validation in the security domain, where contextualizing explanations with program behaviors is critical for trusting the underlying model. Recent research~\cite{warnecke2019don, warnecke2020evaluating, ganz2023hunting} has emphasized verifiable explanations for neural networks in binary analysis and vulnerability discovery, yet \textit{our study is the first to validate explanations within the system provenance domain}.

Explanations have different meanings for model developers and end-users, since end-users cannot make sense of explanations in terms of model weights, but can make sense if explanations are in terms of edges or subgraphs. Therefore, in the security domain, models must explain their decision-making processes~\cite{lipton2018mythos} that are verifiable using the ground truth and domain expertise.

In this paper, we introduce \pname, a novel approach to enhance the explainability of \gnn-based detection models trained on system provenance data by utilizing an interpretable surrogate model, specifically Decision Trees (DT), with security-aware \textit{discriminant subgraph patterns} and \textit{graph structural features}. We efficiently mined subgraph patterns~\cite{jiang2013survey} using a graph evolution miner (GERM~\cite{berlingerio2009mining}) to identify discriminative subgraph patterns between malicious and benign datasets. Our analysis revealed that these subgraph patterns correspond to well-known malicious activities, such as system probing~\cite{mimikatz}, and malware replication~\cite{drakon} and staging~\cite{scar}. Additionally, we extracted structural graph features, such as degree centrality, betweenness centrality, closeness centrality, eigenvector centrality, clustering coefficient, and clustering triangles. These structural features capture long-term dependencies within the system, while the subgraph patterns represent local dependencies. Together, they provide explanations for both short-term and long-term dependencies. 

\pname is specifically designed to address the unique challenges posed by provenance graphs—complex, heterogeneous, and large-scale structures that capture dynamic system behaviors with fine-grained causal semantics. Unlike prior efforts that apply explanation techniques to general-purpose graph domains, our work introduces a domain-specialized explanation framework tailored for provenance-based \gnn intrusion detection. Provenance graphs differ significantly from other graph modalities due to their temporal evolution, multi-typed nodes and edges, and security-relevant semantics embedded in system events. These aspects introduce non-trivial complexities, such as the need to capture temporal causality, reason over multi-hop dependencies, and interpret system-level semantics in a trustworthy manner.

\pname addresses these challenges by adapting DT-based surrogate models with domain informed features, discriminative subgraph patterns and graph-theoretic measures, that are aligned with operational security workflows. For security analysts, \pname offers three critical benefits: \nm{1} it builds confidence that the \gnn bases its predictions on legitimate anomalous subgraphs in the provenance graph; \nm{2} it enables analysts to investigate alerts by localizing explanations to meaningful subgraphs representative; and \nm{3} it equips operators with interpretable justifications to support triage and response decisions.


We emphasize that \pname is an explanation framework, not an attack detection technique. Unlike \gnns, surrogate \dts sacrifice generalizability, expressiveness, and automatic feature extraction to gain interpretability, making them unsuitable for direct attack detection. \pname is a \gnn-structure-agnostic framework capable of incorporating an arbitrary number of subgraph patterns and structural graph features, and it is designed for easy extensibility with additional security-aware graph features.

We focus on the structural features of system provenance graphs, without considering the text or numerical node and edge attributes. We restrict the scope of our attribute view for the following reasons: first, the core merit of provenance-based \ids lies in the ability to retain features drawn from causal dependencies captured in graph structural relations; these structural features are significantly more difficult for adversaries to manipulate than textual attributes (\eg filenames). Secondly, if we cannot explain \gnn decisions based solely on graph structures, it becomes even more challenging to interpret decisions when \gnns incorporate both structural and attribute-based complexities.

We evaluated the effectiveness of \pname using \textit{\fidelityp}/\textit{\fidelitym}, following prior \gnn explainability research~\cite{yuan2020explainability}. An effective explainer should have a higher \fidelityp, as it quantifies the prediction change when important subgraphs are removed, and a lower \fidelitym, as it reflects the prediction change when non-important subgraphs are excluded. We also incorporated two additional measures, \textit{precision} and \textit{recall}, that captures the accuracy of the explanation provided compared against the ground truth. Finally, we introduce a \textit{human actionability} metric that measures the graph traversal distance between an explanation and the ground truth artifacts; a lower score indicates that the explanation guides analysts more directly to the relevant evidence, thereby reducing their investigative workload. In our experimental evaluation, we show \pname surpasses current \sota \gnn explainers in the majority of our security-oriented evaluation tasks.

To demonstrate the advantages of \pname in the system provenance domain, we compared \pname against \sota \gnn explainers such as \gnne~\cite{ying2019gnnexplainer}, \pge~\cite{luo2020parameterized}, and \subx~\cite{yuan2021explainability} across well-balanced security domain tasks, including malware detection, and \apt detection. These \sota \gnn explainers provide the best performance according to other security studies~\cite{ganz2021explaining, warnecke2020evaluating}. The \gnn models in our evaluation are trained with an extensive dataset collected using our in-house deployment which monitored system events on an average of 86 mixed Windows and Linux hosts over 13 months. In addition, in our evaluation, we used the DARPA~\cite{darpa} Transparent Computing dataset, the \apt dataset from~\cite{mukherjee2023sec}, and malware lists from~\cite{survivialism2021sp}.

To the best of our knowledge, our research is the first to leverage security-aware discriminant subgraph pattern and graph structural features specifically tailored for human interpretation of \gnn-based detection models. We summarize the contributions of \pname as follows:

\begin{itemize}[noitemsep,topsep=1pt]
  \item \pname utilizes discriminant subgraph patterns and graph structural features to provide explanation for \gnn-based detection models.

  \item Our security-aware features enabled surrogate \dts to achieve 95\% agreement with the \gnn detection model.

  \item \pname increased \fidelityp (\eg higher is better) by 29\%, precision by 27\%, recall by 25\%, human actionability by 1.4x and decreased \fidelitym (\eg lower is better) by 12\% compared to \sota \gnn explainers.

  \item We curated an extensive dataset of APT scenarios ( \ie DARPA ~\cite{darpa} and past literature~\cite{mukherjee2023sec}) and collected real-world malware samples from ~\cite{survivialism2021sp} to confirm the generalizability of \pname.
\end{itemize}
\section{Background and Related Work}\label{sec:background-and-related-work}


\subsection{System Provenance and Data Collection}

System provenance analysis~\cite{king:2003sosp, inam2023sok} leverages data collection agents on end-hosts to collect interaction events among key system resources: processes, files, and network sockets. This work relies on in-house data from 86 hosts in a university environment, the DARPA Transparent Computing dataset~\cite{darpa}, and datasets from previous study~\cite{mukherjee2023sec}. Our in-house data collection accumulates 13 GB to 92 GB daily, tracking around 875 unique programs, 7,025K processes, 4,824K network connections, and 111,583K file operations.

Our system provenance data schema, detailed in \autoref{tab:schema}, is similar to DARPA's \cdm~\cite{darpa} schema, but we omit memory objects, registry events, and thread distinctions within a process. These choices were made to balance real-world overhead constraints of load balancing and storage. We also established a malicious testbed to generate malware execution traces. To ensure the freshness and realism of our malware samples, we utilize \cti feeds~\cite{attck}, the \code{VirusTotal} database~\cite{vtotal}, and penetration testing \ttp~\cite{metasploit, cobaltstrike, immunity}. We refer to \darpatc dataset~\cite{darpa} and previous literature of \cite{mukherjee2023sec} for \apts.

\subsection{Provenance-based ML Security Solutions}

Recent advancements in system event collection have revitalized host-based \idss. Although initially proposed in the late '90s~\cite{self1996sp}, host-based \idss have proved effective against advanced attacks such as \apts and \fm. Various studies~\cite{wang2020ndss, sigl2021sec, yang2023prographer, zengy2022shadewatcher, cheng2024kairos, rehman2024flash, goyal2024rcaid, bilot12399sometimes, jian2025, wang2024incorporating} demonstrate the efficacy of provenance-based \ml techniques in identifying behavioral deviations. However, the high detection performance is offset by a lack of trustworthy explanations and high false positive rates, impacting operational confidence and resource allocation.
Our work focuses on enhancing the transparency and explainability of provenance-based ML security. We concentrate on a balanced set of security tasks of malware and APT detection. In this paper, we consider graph-level tasks; providing explanations for node and edge level tasks is a promising direction for future work that aligns with upcoming IDS trends.

\heading{\gnns for System Provenance.}
Our research employs an industry-standard \gnn framework (\ie Deep Graph Learning (DGL))~\cite{dgl} to model and explain system provenance graphs. Leveraging DGL's mature development ecosystem \pname aligns with current analytical techniques and streamlined integration into real-world applications. Despite the security community's historical preference for custom detectors~\cite{yang2023prographer, zengy2022shadewatcher, cheng2024kairos, rehman2024flash, goyal2024rcaid, jia2024magic, jian2025, bilot12399sometimes} due to the complexity and heterogeneity of provenance graphs, we chose a general \gnn framework for universal applicability and broader impact. Notably, the DGL community has integrated our heterogeneous \gnn enhancement.
\subsection{Explainability and Security Applications}

\heading{Explaining \ml-based Security Models.}
Due to the importance of explainability in the security domain, several explainability frameworks have been proposed for \ml-based security solutions. LEMNA ~\cite{guo2018lemna} focused on classifying PDF malware and detecting a function's entry point in binary code using regression mixture models as a localized surrogate to approximate the classifier's decision boundary. CFGExplainer~\cite{herath2022cfgexplainer} use deep surrogate models to explain \gnns but work exclusively work on homogeneous graphs.

Jacob \etal~\cite{jacobs2022ai} proposed TRUSTEE, a framework that generates decision tree (\dt)-based interpretations for \ml models to detect shortcut learning. Their success in clarifying model decisions about network packets inspired us to generalize the approach to the system provenance domain. System provenance domain relies on heterogeneous multi-attributed graph datasets, which are not natively consumed by \dts. 

\heading{Explaining \gnn Models.}
Recent research in \gnn explainers \cite{ying2019gnnexplainer, luo2020parameterized, yuan2021explainability} has advanced in identifying key nodes, edges, or subgraphs in \gnns, and are categorized into white-box and black-box explainers. White-box methods, \eg \gnne \cite{ying2019gnnexplainer} and \pge \cite{luo2020parameterized}, access \gnn internals, including model weights and gradients. Conversely, black-box methods like \subx \cite{yuan2021explainability} operate on model inputs and outputs, reducing coupling between the explanation framework and model architecture. ~\cite{kosan2023gnnx} has shown high variance in explanations provided by traditional \gnn explainers, raising reliability and applicability concerns.

\subsection{Ground-truth Verification}\label{sec:documented-entities} 
Ground truth verification of \ml model decisions in security-critical tasks~\cite{warnecke2019don, warnecke2020evaluating, ganz2023hunting} has garnered significant attention, highlighting the role of verifying explanations in unveiling the model's decision-making process. These studies highlight the effectiveness of white-box techniques in malware and vulnerability discovery, and reconstructing ground truth around local explanations \cite{ganz2023hunting}. 

In system provenance, ``ground truth'' refers to the real-world system information (\eg system actions) against which the validity of a model's predictions is checked. In this paper, we approximate the ground truth using documentation created by security vendors, tech reports, and previous studies. The relevant processes, files, and network sockets and their associated events mentioned in the documentation are designated as \textit{documented entities}.

We extract \textit{documented entities} with three methods: \nm{1} we refer to pre-existing malware profile databases that contain information from different security vendors, such as \code{VirusTotal}~\cite{virustotal}, to obtain activity summaries detailing network communications, file system actions, and process behaviors; \nm{2} we extract key entities (\ie process involved, files created and connections made) from tech reports~\cite{darpa:ground, darpa:ground2} and system manuals~\cite{linux:man}; \nm{3} we consult dataset authors~\cite{mukherjee2023sec} and review dataset documentation to identify the components of each attack present in the datasets. We understand that the documentation provided by the security vendors can have experimental errors, as well as selection and experimental bias. To mitigate these problems, we aggregated information from multiple sources. 
\section{Motivation}\label{sec:motivation}

In this section, we present a real-world attack scenario from the DAPRA TC~\cite{darpa:ground2} dataset to highlight the limitations of existing \gnn explainers and demonstrate the effectiveness and utility of \pname. The experimental setup to evaluate \pname is described in \autoref{sec:evaluation} and more case studies are described in \autoref{sec:case-studies} and appendix (\autoref{sec:case3} and \autoref{sec:deatil_attack_desc}).

\begin{figure}[htb]
    \centering
    \begin{subfigure}{0.49\columnwidth}
      \includegraphics[width=\linewidth]{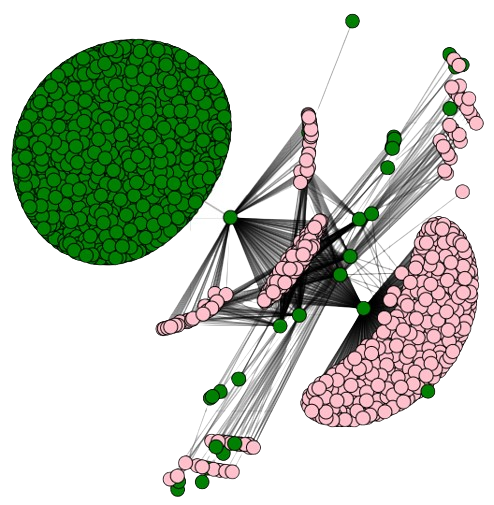}
      \caption{Attack Graph.}
      \label{fig:trace-attack}
    \end{subfigure}
    \hfill %
    \begin{subfigure}{0.49\columnwidth}
      \includegraphics[width=\linewidth]{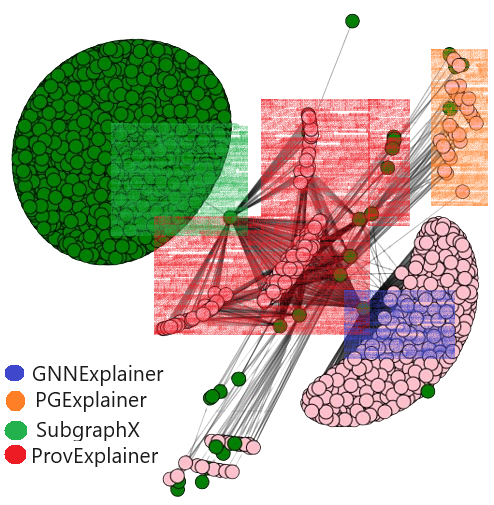}
      \caption{\sota \gnn explainations.}
      \label{fig:trace-sota}
    \end{subfigure}
    \hfill %
    \begin{subfigure}{0.90\columnwidth}
      \includegraphics[width=\linewidth]{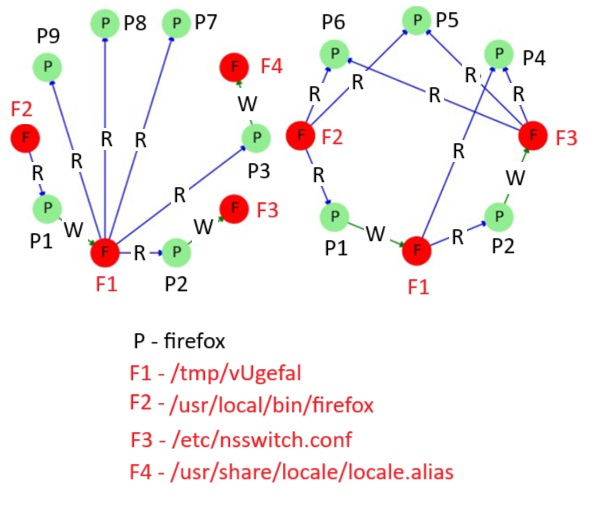}
      \caption{Pattern \#1 and Pattern \#2.}
      \label{fig:trace_p2}
    \end{subfigure}
    \caption{Trace: after an employee clicks on a phishing link, \code{Firefox} installs multiple Trojans to exfiltrate sensitive data.}
    \label{fig:overall_trace}
  \end{figure}
  
  \heading{Attack Scenario (Trace: Phishing Email~\cite{darpa}).}\label{sec:case1}
  \autoref{fig:overall_trace} presents a simplified depiction of the APT attack initialized by phishing e-mail link. This particular attack was executed as part of a DAPRA red team exercise by Trace team.  
  First, the attacker launches a phishing campaign that compromises the identity of an employee. Leveraging this compromised identity, the attacker then sends deceptive emails containing links to a malicious website (\ie \url{www.nasa.ng}) to other employees. When an employee visits the website, a Trojan is downloaded and placed in \path{/tmp/vUgefal}. Upon execution, the Trojan replicates itself within the system by reading from its own file. 

  The malware adopts the name \code{firefox nasa.ng} and stores sensitive content in \path{/usr/share/locale/locale.alias} for exfiltration. To evade detection, the Trojan creates additional malicious copies using benign program names (e.g., \code{firefox}), allowing it to replicate unhindered and overwhelm the system with malicious processes. These malware clones quietly read sensitive user files while the original Trojan remains undetected.

  \heading{Limitation of \sota \gnn explainers.}
  \autoref{fig:trace-sota} compares the explanations produced by \pname with those of \sota \gnn explainers (\ie \gnne, \pge, and \subx) in capturing malicious activities relevant to \gnn-based \ids. While \gnne effectively uncovers the initial compromise behavior such as payload dropping and staging, it fails to capture later attack stages notable the interactions with shared libraries and cache files. \pge, in contrast, highlights irrelevant benign system components like \code{xfce4-appfinder} (invoked by a DARPA script to simulate an enterprise environment) but overlooks critical malicious interactions. \subx accurately detects the malicious inheritance behavior of \code{Firefox} creating multiple malware clones, yet it misses key points such as how the malware writes its payload to disk and accesses sensitive files. 

  \pname presents a complete view of the kill chain from Trojan deployment and clone creation to the final step of accessing sensitive data. \pname uses security relevant discriminative subgraph patterns and graph structural metrics as features to capture the holistic malicious behavior within the system. Moreover, it accomplishes this despite functioning as a black-box solution, requiring neither direct access to the underlying model nor the training data. This flexibility makes ProvExplainer compatible with a wide range of GNN-based IDS deployments.

  \heading{\pname Features:} Clustering coefficient (graph structural metric); pattern \#1 and \#2 (discriminative subgraph patterns).
  
  \heading{System Interpretation of \pname Features.}
  In this \apt scenario, clustering coefficient, pattern \#1 and pattern \#2 in \autoref{fig:trace_p2} showcase the malware's structures. Pattern \#1 reveals the Trojan's propagating activity, where initial malware (Process P1) reads the firefox binary (\code{/usr/local/bin/firefox}) to execute it and write the maliclious payload (\code{/tmp/vUgefal}) to a file that is read by other malicious \code{firefox} processes (Process P4, P7,P8, P9). A series of intra-malware communication happens when the another malicious \code{firefox} (Process P2) reads the malicious payload and writes to system file ( \code{/etc/nsswitch.conf}). As seen in pattern \#2, other malware processes (Process P4, P5, P6) read the system file to update their configurations and task lists to better attack the victim. This propagation of information continues such that any subsequent malware processes that read from the system file gets the updated configuration.
  
  By masquerading as benign programs, the malware disguises its malicious processes, allowing it to proliferate undetected. The clustering coefficient of the graph increases because of multiple malware reading the template files by forming large process creation and file read clusters. 
\section{Problem Statement and Threat Model}\label{sec:threat-model}
Our research addresses explainability in \gnn-based security models built on system provenance graphs, tackling a core issue in the security domain. The complexity of explaining \gnn decisions is exacerbated by graph structural learning, which adds to the inherent complexity of \nns. Existing studies on \gnn explainability~\cite{yuan2020explainability, gnnexplainer, subgraphx, pgexplainer, herath2022cfgexplainer, ganz2023hunting} often fail to effectively map back to system behaviors in the provenance domain. To bridge this gap, \pname, employs a surrogate \dt-based method that utilizes interpretable security-aware graph structural features (\ie discriminant subgraph patterns and graph structural features) to explain \gnn decisions.  Given a graph $\mathcal{G} = (\mathcal{V}, \mathcal{E})$ and a \gnn model $\mathcal{M}$, an explanation method $\mathcal{EM}$ produces:
\begin{itemize}
    \item Set of subgraphs $\mathcal{SG}$, where $\mathcal{SG} = \{ \mathcal{G}_1, \mathcal{G}_2, \dots, \mathcal{G}_k \}$ and $\mathcal{G}_i = (\mathcal{V}_i, \mathcal{E}_i)$ for $i = 1..k$, with $\mathcal{V}_i \subseteq \mathcal{V}$ and $\mathcal{E}_i \subseteq \mathcal{E}$.
    \item Set of properties $\mathcal{P}rop = \{ p_1, p_2, \dots, p_j \}$, where each $p_j$ is a graph structural property of $\mathcal{G}$.
\end{itemize}
The subgraphs $\mathcal{SG}$ and properties $\mathcal{P}rop$ are designed to highlight the most relevant components of $\mathcal{G}$ that contribute to the predictions of $\mathcal{M}$.

Our threat model assumes the integrity of on-device data collection, relying on provenance records secured by existing mechanisms (\cite{provdetector, sigl2021sec, nodoze, mukherjee2023sec, liu2018ndss,cheng2024kairos, rehman2024flash, goyal2024rcaid}).
Our primary objective is to generate security-aware explanations to aid security analyst and increase their trust in the \gnn's decisions. We consider graph-level anomaly detection tasks; explaining \gnn decisions in node/edge level tasks is outside the scope of this work. Systematically aggregating an accurate and trustworthy ground truth for malware and APT behaviors poses a challenging open problem. In this paper, we approximate the ground truth using publicly available documentation (\autoref{sec:documented-entities}). In line with recent literature on \gnn{} explanation~\cite{herath2022cfgexplainer,warnecke2019don, warnecke2020evaluating}, adversarial samples are outside the scope of the paper. Creating robust detection and explanation systems that can withstand adversarial manipulation~\cite{goyal2023sometimes, mukherjee2023sec}, dataset poisoning, and detection model manipulations are critical open research problems that are orthogonal to our work.

\section{\pname}\label{sec:system-overview}

\begin{figure}[tb]
    \includegraphics[width=1.0\linewidth]{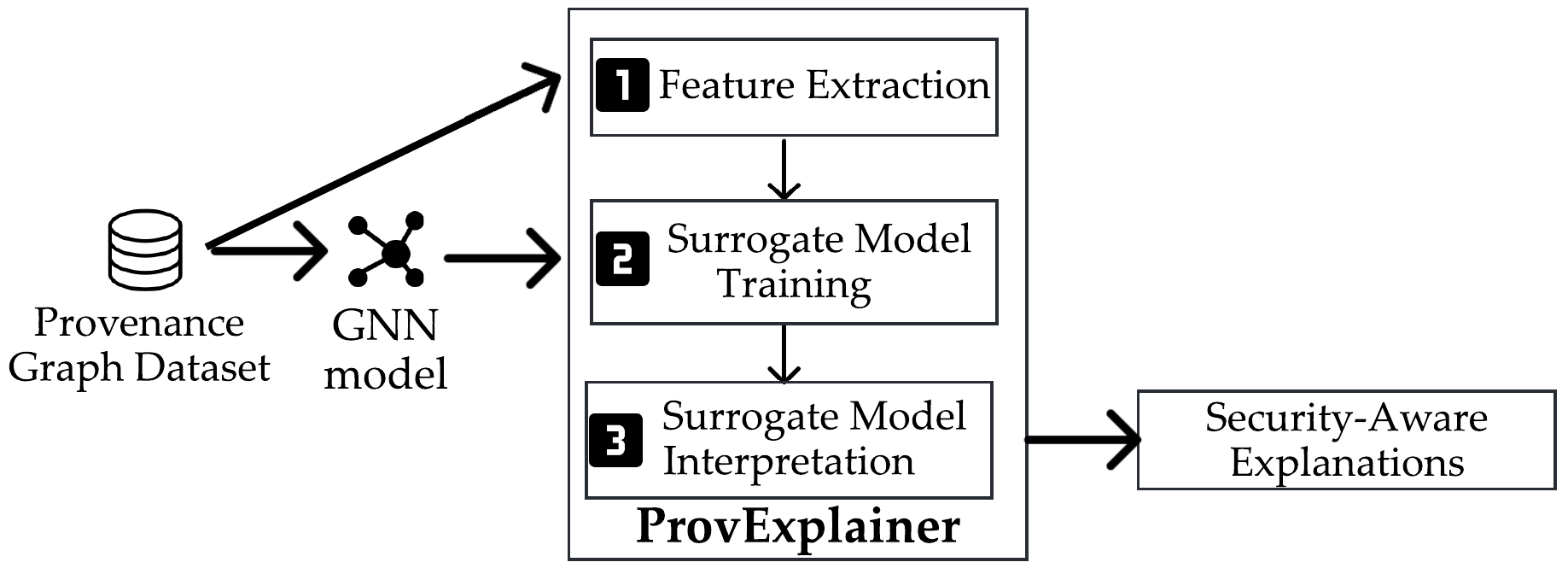}
    \caption{\pname architecture.}
    \label{fig:provexplinerfig}
\end{figure}

We apply \pname in three stages (\autoref{fig:provexplinerfig}).

\heading{Stage 1: Extract Security-aware Features (\autoref{sec:sec_domain_feats}).} 
We extract security-relevant features such as subgraph patterns using a discriminative subgraph miner~\cite{jiang2013survey} and graph structural features such as clustering and centrality measures that facilitate distinction between benign and anomalous datasets. The subgraph patterns enable the identification of localized attack vectors, while global graph structural features capture holistic attack behaviors.

\heading{Stage 2: Train an Interpretable Surrogate Model (\autoref{sec:extracting-decision-trees}).}
Next, we utilize diverse system provenance dataset to train an interpretable surrogate \dt that aligns with the \gnn decisions using the extracted features. Since the \gnn's decision-making process relies on a structure-dependent message-passing algorithm, we efficiently mimic the decision-making process of the \gnn's using the security-relevant features.

\heading{Stage 3: Interpret the Surrogate Model (\ref{sec:interpret_model}).}
To interpret the explanation for detection using the surrogate \dt, we extract the graph nodes participating in the subgraph patterns that contribute to the \dt's decision. These explanations are valid only when the surrogate agrees with the \gnn.

\subsection{Feature Extraction}\label{sec:sec_domain_feats}

\begin{algorithm}[h]
    \small
    \caption{Discriminative Pattern Mining}
    \label{alg:disc}
    \begin{algorithmic}[1]
        \Require Benign $D_b$ and Anomaly $D_a$ datasets, $k$, and $score\_func$
        \Ensure Top-$k$ Discriminative Subgraph Patterns
        \State $Patterns \gets$ [ ]
        \State $CP \gets \text{GERM}(D_a, k, \text{minSupport}=1)$
        \For{each pattern $p \in CP$}
            \State $div \gets \text{getDIVERSITY}(p)$
            \For{each graph $G \in D_b$}
                \State $size_b, support_b \gets \text{getSIZEandSUPPORT}(p, G)$
            \EndFor
            \State $cov_b \gets \text{getCOVERAGE}(p, D_b)$
            \State $score_b \gets score\_func(div, cov_b, size_b, support_b)$

            \For{each graph $G \in D_a$}
                \State $size_a, support_a \gets \text{getSIZEandSUPPORT}(p, G)$
            \EndFor
            \State $cov_a \gets \text{getCOVERAGE}(p, D_a)$
            \State $score_a \gets score\_func(div, cov_a, size_a, support_a)$
            \State $Patterns.\text{append}([p, |\text{score}_a - \text{score}_b|])$
        \EndFor

        \State Sort $Patterns$ in desc order based on the score difference
        \State \Return $Patterns[:k]$
    \end{algorithmic}
\end{algorithm}

\heading{Security-relevant Features.}
The extraction of security-relevant subgraph features is pivotal in distinguishing anomalous patterns from benign patterns within provenance datasets. To avoid information leakage and ensure no data snooping, \textit{patterns that are mined exclusively on the training dataset} are used for explanation evaluation. This approach enables unbiased pattern discovery and allows the identified discriminative features to be evaluated solely on the testing dataset, maintaining a rigorous evaluation protocol.

Frequent subgraph patterns are mined from the training dataset using a graph evolution miner (GERM~\cite{berlingerio2009mining}), where subgraphs of specified \emph{sizes} and their \emph{supports} are identified. Support means how many times the subgraph instances were identified in the graph. These patterns serve as representative features, capturing the frequency and structure of motifs present in the datasets.

For each subgraph pattern, we calculate its \emph{pattern diversity}, which measures the structural complexity of the pattern. Diversity scores are computed by aggregating node and edge weights, where node weights are assigned according to unique node labels, and edge weights reflect the types of connections between nodes. This diversity score quantifies the variability within each pattern, aiding in highlighting complex, potentially discriminative structures.

We also measure the \emph{coverage} of a subgraph pattern by measuring its presence across graphs within the dataset, providing a statistical representation of how commonly the pattern occurs in the dataset as a whole. Calculated as the proportion of graphs in which the pattern appears with non-zero support, coverage serves as an indicator of the pattern's prevalence or rarity in benign and anomalous instances. High coverage patterns suggest commonly occurring structures within the respective dataset, representing anomalous or benign operations. We want to identify patterns that have high coverage in anomalous but low coverage in benign datasets, providing a basis for distinguishing between them.  

To quantify the discriminative utility of each pattern, we define a \emph{scoring function} parameterized by weights $\alpha, \beta, \gamma, \text{ and } \eta$, which represent the relative importance of diversity, coverage, pattern size, and support, respectively. The $\text{score\_func}$ in \autoref{alg:disc} (\ie scoring function) is:
\[
\text{Score} = \alpha \cdot \code{div} + \beta \cdot \code{cov} + \gamma \cdot \code{size} + \eta \cdot \code{support}
\]
where \code{div} denotes the diversity of the pattern, \code{cov} represents the proportion of graphs in which the pattern appears, \code{size} is the average number of nodes in the pattern, and \code{support} reflects the frequency of pattern occurrence within each graph. 

Increasing $\alpha$ enhances the selection of patterns with diverse node types, which may improve the ability to identify unique patterns but could reduce the focus on more frequent patterns. Raising $\beta$ results in higher scores for patterns that appear across a larger number of graphs, promoting generalization but potentially overlooking rare but important patterns. Elevating $\gamma$ favors patterns with larger node counts, which can capture more complex interactions but might exclude simpler yet crucial substructures. Finally, increasing $\eta$ prioritizes patterns with higher frequency, amplifying commonly occurring patterns but at the expense of pattern diversity. 

For our study, we used the following weights: $\alpha = 0.2$, $\beta = 0.4$, $\gamma = 0.1$, and $\eta = 0.3$. These values were selected empirically to prioritize coverage and support as the primary differentiators, while also ensuring that the extracted patterns are diverse and large.
Subgraph patterns are extracted from each graph in the dataset using \autoref{alg:disc} but the discriminant subgraph patterns are selected exclusively from the anomalous dataset, as we ensure that patterns contributing significantly to the anomalous dataset's structure are prioritized. Patterns that maximize the score difference between benign and anomaly datasets are selected as discriminative patterns, as they effectively differentiate between benign and anomalous behaviors.  Also, subgraph patterns are extracted solely from the training set and evaluated on an independent test set, thereby eliminating any risk of data leakage.


\heading{Graph Structural Features.}\label{sec:graph-structural-features}
From graph theory literature, we selected the following structural features~\cite{zaki2014data}: degree centrality, closeness centrality, betweenness centrality, eigenvector centrality, clustering coefficient, and clustering triangles. These standard features are known for capturing the key characteristics of graphs, serving as a custom signature that effectively reflects the global properties of a graph. 
To refresh the reader, we provide informal definitions and their meaning in system provenance domain.     

\headinggi{Degree Centrality.}
Degree centrality is simply the degree of the node, or the count of its incident edges. The degree centrality of a node indicates the extent of its direct connections in the graph. When aggregated across all nodes, high degree centrality values suggest a densely connected system with significant interactivity, whereas low values indicate sparsely connected regions, potentially reflecting isolated or specialized nodes. High degree centrality could suggest the presence of core process hubs, while low values imply a more dispersed system with no central process hubs. However, degree centrality alone cannot definitively indicate malicious behavior. For example, a long-running attack creating a long process chain shows a low degree of centrality, while attacks that interact with many files or fork multiple child malware processes will exhibit a high degree of centrality. In the case of a cryptominer masquerading as \code{schtasks.exe}, we expect the benign \code{schtasks.exe} to show low centrality as it schedules tasks, while the cryptominer malware have high centrality due to spawning multiple malicious processes.

\headinggi{Closeness Centrality.}
Closeness centrality is the inverse of the mean shortest path distance from all other nodes to the target node, indicating how close the nodes are within the graph. Closeness centrality measures how centrally located a node is within the graph by examining its proximity to other nodes. High closeness centrality across the graph suggests that many nodes are relatively close to each other, indicating efficient communication pathways. In contrast, low closeness centrality values highlight dispersed sections of the graph, potentially representing decentralized or peripheral interactions within the system. High closeness centrality indicates well-connected process hubs central to the system, while low values suggest that processes are widely dispersed with limited central hubs. Empirically, system processes often have high closeness centrality, while user-level or malware processes show lower values. For instance, botnets masquerading as system-level processes will be distinguishable, since botnet will have a higher closeness centrality due to multiple network connecting being created due to their DDoS activities.

\headinggi{Betweenness Centrality.}
Betweenness centrality measures the frequency with which a node appears on the shortest paths between other nodes, highlighting its role as a waypoint. It indicates how often a node acts as a bridge along the shortest paths between other nodes. High betweenness centrality suggests a system's reliance on certain core hubs or bottleneck paths, while lower values indicate multiple alternative pathways and a more distributed structure. High betweenness centrality indicate processes that serve as intermediaries in data or control flow, whereas low values suggest a network of distributed pathways with minimal reliance on specific process hubs. Malware like worms and spyware tend to have low betweenness centrality, as they seek to spread across the system and create multiple distributed information pathways.

\headinggi{Eigenvector Centrality.}
Eigenvector centrality assigns higher values to nodes connected to other influential nodes, reflecting the prominence of node clusters that are influential. High eigenvector centrality in processes indicates hubs with extensive control over information flow, while low values imply a flatter structure without prominent hubs. Ransomware have high eigenvector centrality, as malicious processes involved in reading and encrypting files act as critical nodes. They keep few child processes to allow high-volume file encryption without overloading the system.

\headinggi{Clustering Triangles.}
Clustering triangles count the number of triangles that include the node. The clustering triangles count represents the frequency of triangular connections in the graph. High clustering triangle values indicate localized connectivity, suggesting shared resources or close interactions among nodes, while low values indicate a more hierarchical or linear flow. Malware often creates prominent triangles when they write template files and repeatedly execute them, making copies of themselves. Similarly, legitimate programs like \code{svchost.exe} (a service scheduler) form triangles when creating \code{conhost.exe} and reading standard \code{*.dll} files. In contrast, ransomware reads files to encrypt them, creating ``spoke-and-wheel'' shape rather than triangles, resulting in lower triangle counts than benign processes, \code{svchost.exe}.

\headinggi{Clustering Coefficient.}
The clustering coefficient is calculated by dividing the number of triangles the node participates in by its degree, indicating how close its connections form a complete graph with its neighbors. The clustering coefficient measures the degree to which nodes’ neighbors are interconnected. A high clustering coefficient suggests that nodes form close-knit groups, often indicating shared resources or community-like interactions, while a low coefficient reflects more dispersed connections. High clustering coefficients point to strong groupings around key processes with shared resources, while low values suggest processes with less close-knit hubs. The clustering coefficient provides a similar interpretation to closeness centrality; however, while closeness centrality highlights a process's accessibility within the system, the clustering coefficient emphasizes the density of its immediate neighborhood connections.

\subsection{Surrogate Model Training}\label{sec:extracting-decision-trees}
To obtain explanations from our security-aware graph structural features, we use them to train a global surrogate \dt. By training a \dt to agree with the predictions of a \gnn model, we can interpret the \dt to gain insights about the \gnn's decision-making process. To achieve the best agreement results, we enhance traditional \dt training with data augmentation that iteratively increases the weight of incorrectly classified samples~\cite{trustee2022ccs}.

We begin with a labeled set of graphs $D_G=(G, Y)$, which is used to train the \gnn. \gnn's predictions on $D_G$ are collected, yielding $GNN(D_G) = Y'$. To prepare the dataset for training the \dt, we extract the discriminative subgraph patterns and graph structural features (\autoref{sec:sec_domain_feats}) $F$ and associate them with the \gnn's predictions $Y'$ to create a labelled feature dataset $D_F=(F, Y')$, which we split into train, validation, and testing to evaluate the surrogate \dt.


\subsection{Surrogate Model Interpretation}\label{sec:interpret_model}

We obtain instance-level explanations after training the surrogate \dt. Each decision node in the \dt corresponds either to a discriminant subgraph pattern or to a graph-structural measure (with its associated threshold). Nodes closer to the root of the tree exert greater influence on the decision path; therefore, we rank subgraph-participating nodes based on their relative position in the \dt.  

In cases where \nm{1} the surrogate model's prediction diverges from that of the \gnn, and \nm{2} the decision path is composed solely of structural measures without subgraph patterns, node-level explanations cannot be extracted, even though a \dt is generated. While structural metrics alone cannot precisely localize specific nodes, they remain crucial indicators of long-running, persistent \apt campaigns. This is because during an \apt attack, adversaries may preserve node-level attributes, but they cannot maintain the benign program's graph structure.  

Finally, the interpretable surrogate \dt enables the construction of actionable, security-aware explanations for individual predictions. Using \pname's global surrogate \dt and participating node's system attributes (\eg process names, file names, and IP/PORT), domain experts can leverage this enriched contextual information to understand the decision-making process of the \gnn.

\section{Evaluation}\label{sec:evaluation}

In this section, we evaluate \pname{}'s effectiveness in explaining stealthy attacks. We aim to answer the following research questions (RQs):

\begin{itemize}[noitemsep, leftmargin=0.9cm]
    \item[{\bf RQ1:}] {\bf Explanation Accuracy.} Can \pname{} accurately explain APT and \fm detections (\autoref{sec:graph_structure_eval} and \autoref{sec:ablation_gat})?
    \item[{\bf RQ2:}] {\bf Comparison with \sota \gnn Explainers.} How do \pname's explanations compare against \sota \gnn explainers (\eg \gnne\cite{ying2019gnnexplainer}, \pge\cite{luo2020parameterized}, and \subx\cite{yuan2021explainability}) (\autoref{sec:comp_sota})?
    \item[{\bf RQ3:}] {\bf Explainability Effectiveness.} How much actionable are \pname's explanations  (\autoref{sec:human_study})?
\end{itemize}

\subsection{Evaluation Protocols}\label{sec:eval_design}
We evaluate \pname on a different \apt detection and \fm datasets: \nm{1} publicly available APT attack simulations~\cite{darpa}, \nm{2} an APT dataset from a recent study~\cite{mukherjee2023sec}, and \nm{3} execution traces of Fileless Malware~\cite{survivialism2021sp}. We implement two general-purpose \sota \gnn-based detector models, \gat~\cite{velivckovic2017graph} and \sage~\cite{hamilton2017inductive}, following the methodology adopted in recent explanation literature~\cite{herath2022cfgexplainer, kosan2023gnnx}. Our focus is to first thoroughly understand and explain \ids that directly leverage \gnns for security-critical tasks, before addressing more complex \ids architectures that use \gnns to extract graph embeddings for downstream anomaly detection, as discussed in~\cite{bilot12399sometimes}. Accordingly, we do not compare against recent \gnn-based \ids~\cite{zengy2022shadewatcher, rehman2024flash, jia2024magic, yang2023prographer, cheng2024kairos, goyal2024rcaid, jian2025, bilot12399sometimes} that rely on custom graph embeddings and employ a separate model for final detection.

We conducted an ablation study to understand the impact of different feature sets (\eg subgraph patterns and graph structural features). We also evaluate the explanations given by \pname{} against \sota{} \gnn{} explainers~\cite{ying2019gnnexplainer, luo2020parameterized, yuan2021explainability}. Finally, we conduct a non-trivial human-centered evaluation to validate the interpretability of the explanations from a security-analyst perspective. In \autoref{sec:case-studies}, we present case studies explaining \apt detection and offer in-depth insights into how \pname provide security-relevant explanations.

\subsection{Evaluation Metrics.}\label{sec:eval-metric}
In our evaluation of \pname, we focus on three critical aspects. The first is the agreement of the surrogate \dts with the \gnn model, which we measure using the weighted macro averaged (WMA) F1 score of the surrogate \dt's predictions with respect to the \gnn's predictions. The \textit{agreement} metric gauges the faithfulness of the \dt in replicating the conclusions of \gnns. WMA F1 score helps to account for the data imbalance issue present in the anomaly detection datasets.

The second metric is \fidelityp and \fidelitym which is used by different \gnn explainers~\cite{yuan2020explainability}, which measure if removing the important subgraph pattern causes prediction change (\fidelityp), and if removing unimportant subgraph patterns ensures prediction remains unchanged (\fidelitym). Let a GNN explanation method $\mathcal{EM}$ explain a prediction $y$ made by a GNN model $\mathcal{M}$ for a graph $G=(\mathcal{V}, \mathcal{E})$. Let $\mathcal{EM}$ output a set of important nodes $\mathcal{V}_{imp}$ (\ie top $k$ nodes that participate in important edges or subgraphs and $k$ be the number of nodes considered). Therefore, \fidelityp $=\frac{1}{N} \sum_{i=1}^{N} \big(\mathds{1}(\hat{y}_i=y_i) - \mathds{1}(\hat{y}_i^{\text{imp}}=y_i)\big)$, where $N$ is the number of graphs, $y = \mathcal{M}(\mathcal{G})$, $y^{\text{imp}} = \mathcal{M}(\mathcal{\tilde{G}})$, and $\mathcal{\tilde{G}}$ is $G$ after removing the top $k$ important nodes. Similarly, \fidelitym $=\frac{1}{N} \sum_{i=1}^{N} \big(\mathds{1}(\hat{y}_i=y_i) - \mathds{1}(\hat{y}_i^{\text{unimp}}=y_i)\big)$, where $y^{\text{unimp}} = \mathcal{M}(\mathcal{\hat{G}})$ and $\mathcal{\hat{G}}$ is $G$ after removing a random set of $k$ unimportant nodes, where unimportant nodes are nodes that do not participate in critical subgraphs. \textit{A good explainer is the one that will have a high \fidelityp and a low \fidelitym.} Since, after removing important (relevant) edges the prediction $y$ should change and similarly, after removing unimportant edges the prediction $y$ should not change.

To evaluate \pname's and \sota \gnn explainers' proficiency in identifying security-relevant entities, we define precision and recall metrics with respect to documented entities \autoref{sec:documented-entities}. Let $\mathcal{D}$ be the set of documented entities and using the above definitions for $\mathcal{V}_i$ and $k$. \textit{Precision} is the proportion of explanation nodes that are documented, and \textit{recall} is the fraction of documented entities retrieved: $precision(V_k, k, \mathcal{D})=\frac{|V_k \cap \mathcal{D}|}{k}$, and $recall(V_k, \mathcal{D})=\frac{|V_k \cap \mathcal{D}|}{|\mathcal{D}|}$.

Lastly, to assess the actionability of the explanations from a security analyst's perspective, we introduce a graph traversal distance metric, where we compute the shortest-path distance between each node recommended by an explainer and its corresponding ground-truth node (\autoref{sec:documented-entities}). Let $\mathcal{V}_\text{GT}$ denote the set of ground-truth nodes and $\mathcal{V}_\text{expl}$ the set of source nodes highlighted by the explainer. The distance metric is defined as: $distance_{traversal} = \frac{1}{|\mathcal{V}_\text{GT}|} \sum_{i=1}^{|\mathcal{V}_\text{GT}|} d(v^\text{expl}_i, v^\text{GT}_i)$, where $d(v^\text{expl}_i, v^\text{GT}_i)$ is the shortest-path node distance between the $i$-th recommended node and its matched ground-truth node in the graph. This node-based distance is then normalized by the number of ground-truth nodes $|\mathcal{V}_\text{GT}|$, ensuring a consistent scale across graphs. The intuition behind this metric is that a lower distance indicates the explainer's recommendation is closer to the true malicious entities, thereby reducing the analyst's cognitive and temporal burden during an investigation.

\begin{figure*}[h]
  \centering
  \resizebox{\textwidth}{!}{%
    \begin{minipage}{\textwidth}
      \centering
      \begin{minipage}[t]{0.49\textwidth}
        \vspace{0pt}
        \centering
        \begin{subfigure}[t]{0.99\linewidth}
          \centering
          \includegraphics[page=1,width=\linewidth]{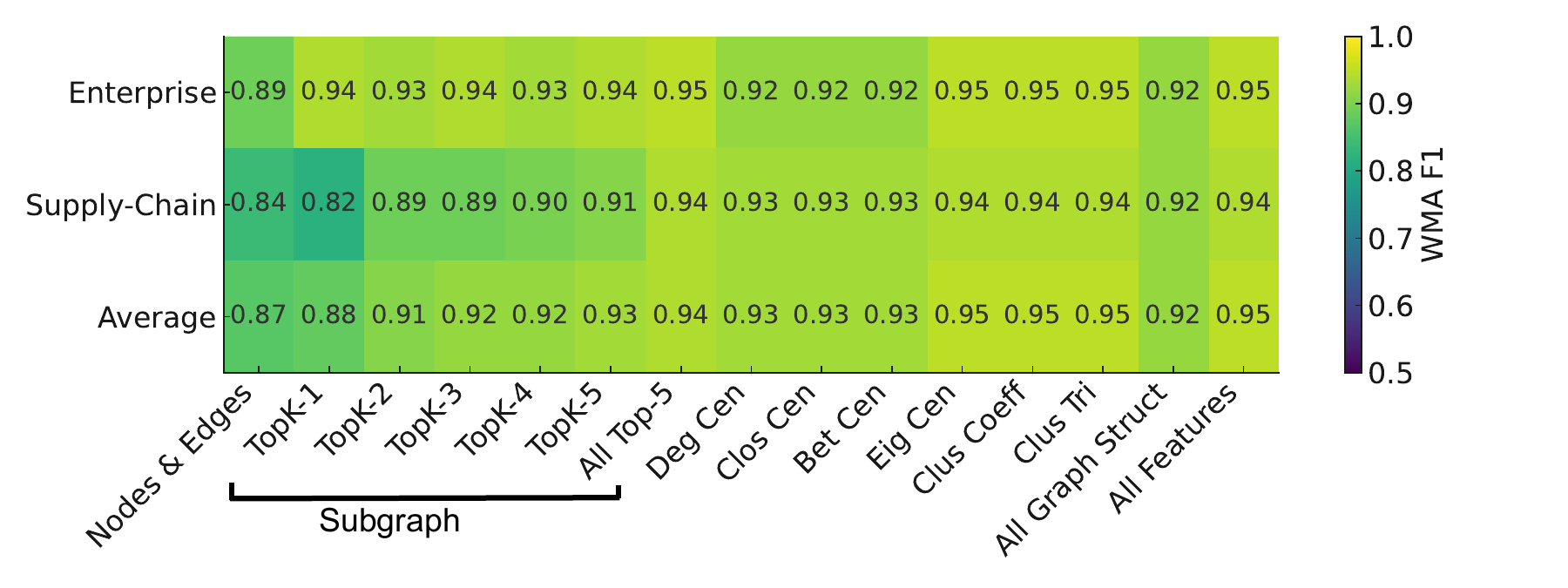}
          \caption{DARPA APT Dataset~\cite{darpa}}
          \label{fig:gat_wma_f1_heatmap_mukh}
        \end{subfigure}

        \begin{subfigure}[t]{0.99\linewidth}
          \centering
          \includegraphics[page=2,width=\linewidth]{figs/gat_wma_f1_graphs.pdf}
          \caption{APT Dataset from \cite{mukherjee2023sec}}
          \label{fig:gat_wma_f1_heatmap_darpa}
        \end{subfigure}
      \end{minipage}
      \hfill
      \begin{minipage}[t]{0.49\textwidth}
        \vspace{0pt}
        \centering
        \begin{subfigure}[t]{0.99\linewidth}
          \centering
          \includegraphics[page=3,width=\linewidth]{figs/gat_wma_f1_graphs.pdf}
          \caption{Fileless Malware from \cite{survivialism2021sp}}
          \label{fig:gat_wma_f1_heatmap_fileless}
        \end{subfigure}
      \end{minipage}
    \end{minipage}%
  }
  \caption{Agreement (higher is better) measured using WMA F1 of surrogate decision trees versus the \gat model across different feature sets.}
  \label{fig:gat_f1}
\end{figure*}

\subsection{Evaluation Tasks}

\heading{APT Detection.}
We utilized the DARPA Transparent Computing (TC) Data Releases~\cite{darpa} and previous literature~\cite{mukherjee2023sec} for \apt attack detection. This dataset, encompassing various OSes, provides a comprehensive basis for advanced security research. The DARPA \apts were designed to attack a system which consists of long-running processes and captured the stealthy attack vectors frequently employed by advanced adversaries. We particularly focused on three DARPA datasets used by previous studies~\cite{zengy2022shadewatcher, cheng2024kairos, rehman2024flash, goyal2024rcaid}: FiveDirections, Trace, and Theia. However, these tasks were conducted in a simulated environment and only lasted for two weeks, involving a limited number of hosts. Therefore, we also evaluated against the \apt scenarios (\eg Enterprise APT and Supply-Chain APT) conducted by ~\cite{mukherjee2023sec}, which were performed with realistic benign background workloads. Interested readers can find detailed statistics presented in \autoref{table:anomaly_vertex_edge}.

\heading{Fileless Malware Detection.}
For Fileless Malware detection, we targeted a family of stealthy malware samples that impersonate benign programs, evading conventional security solutions but are detectable by \gnn-based provenance analysis. The malware samples were chosen in accordance with guidelines from the literature~\cite{kuchler2021does, avllazagaj2021malware} to minimize experimental bias and ensure freshness. The Fileless Malware dataset includes various categories~\cite{kuchler2021does}, including banking Trojans, spyware, ransomware, and malware installers; detailed statistics presented in \autoref{table:anomaly_vertex_edge} in the appendix.

The Banking banking trojan~\cite{kuchler2021does} steals banking credentials from victim machines and spreads through spam and compromised download links. To camouflage interactions with \code{.dlls} and temporary files, it masquerades as \code{svchost.exe}. The Spyware, Ulise~\cite{kuchler2021does}, is a multi-purpose Trojan that can establish remote access connections, capture keyboard input, collect system information, upload files, drop other malware into the infected system, and perform encryption. Because it propagates through the network sockets and interacts with many system files, it masquerades as \code{python.exe} as it needs a target that exhibits diverse behavioral patterns.

\begin{table}[H]
    
    \centering
    \resizebox{0.85\columnwidth}{!}{%
        \begin{tabular}{lcccc}\toprule
            Dataset & \makecell{\gat} & \makecell{\trusteedt \\ (agree w/\ GAT)} & \makecell{\sage} & \makecell{\trusteedt \\ (agree w/\ GraphSAGE)} \\
            \cmidrule(lr){1-1}
            \cmidrule(lr){2-3}
            \cmidrule(lr){4-5}
            \multicolumn{5}{c}{DARPA APT Dataset~\cite{darpa}} \\
            \midrule
            FiveDirections   & 0.82 & 0.88 & 0.82 & 0.85 \\
            Trace            & 0.93 & 0.93 & 0.87 & 0.87 \\
            Theia            & 1.00 & 1.00 & 0.94 & 0.94 \\
            \midrule
            \textbf{Average} & 0.92 & 0.93 & 0.88 & 0.89 \\
            \midrule
            \multicolumn{5}{c}{APT Dataset from \cite{mukherjee2023sec}} \\
            \midrule
            Enterprise       & 0.82 & 0.95 & 0.81 & 0.94 \\
            Supply-Chain     & 0.82 & 0.94 & 0.80 & 0.97 \\
            \midrule
            \textbf{Average} & 0.82 & 0.95 & 0.81 & 0.95 \\
            \midrule
            \multicolumn{5}{c}{Fileless Malware from \cite{survivialism2021sp}} \\
            \midrule
            \code{wscript.exe}  & 1.00 & 1.00 & 0.95 & 0.95 \\
            \code{cscript.exe}  & 1.00 & 1.00 & 1.00 & 1.00 \\
            \code{reg.exe}      & 1.00 & 1.00 & 1.00 & 1.00 \\
            \code{python.exe}   & 0.99 & 0.99 & 0.99 & 0.99 \\
            \code{explorer.exe} & 0.98 & 0.98 & 1.00 & 0.98 \\
            \code{netsh.exe}    & 0.98 & 0.98 & 0.98 & 0.98 \\
            \code{net.exe}      & 0.97 & 0.97 & 0.97 & 0.97 \\
            \code{rundll32.exe} & 0.96 & 0.96 & 0.96 & 0.96 \\
            \code{mshta.exe}    & 0.95 & 0.95 & 0.85 & 0.85 \\
            \code{schtasks.exe} & 0.93 & 1.00 & 0.93 & 1.00 \\
            \code{conhost.exe}  & 0.89 & 0.89 & 0.93 & 0.93 \\
            \code{svchost.exe}  & 0.86 & 1.00 & 0.86 & 1.00 \\
            \midrule
            \textbf{Average}   & 0.96 & 0.98 & 0.95 & 0.97 \\
            \bottomrule
        \end{tabular}
    }
    \caption{{Surrogate \dts have high agreement ($\geq$ 0.85) with the decisions of \gnn measured using the WMA F1 score.}}
    \label{table:dt_accuracy_fidelity}
\end{table}

\subsection{Graph Structural Feature Evaluation}\label{sec:graph_structure_eval}
To answer \textbf{RQ1}, \autoref{table:dt_accuracy_fidelity} demonstrates the effectiveness of surrogate \dts in mirroring the decision process of \gnn models like \gat and \sage. Agreement between the surrogate \dt and the \gnn is an important metric for two reasons: \nm{1} surrogate explanations are only valid when the surrogate agrees with the \gnn, and \nm{2} high agreement indicates that the surrogate model is a good approximation of the \gnn's decision-making process.

\pname exhibits the highest agreement($>$ 97\%) with the \gnn models on the \fm datasets, but the \apt datasets only showed good agreement ($>$ 93\%). The surrogate \dts' superior performance in \fm datasets can be attributed to \pname's ability to capture local subgraph patterns with high efficiency but \apt attacks that are long running and utilize evasive tactics (noted by previous studies~\cite{mukherjee2023sec,survivialism2021sp}) are harder to detect and explain. The dataset such as FiveDirections and \code{conhost.exe} are dominated by attackers using stealthy techniques such as \emph{living-off-the-land}, which involve memory object interactions that are currently not captured in our provenance graphs.

The absence of the distinguishing features or rare features that are exclusively present in the testing dataset impairs the effectiveness of \pname. We observed that \gat outperforms \sage in majority of the datasets. This trend is attributed to \gat's ability to assign varying importance to different neighborhood structures and capture long-range structural dependencies through its attention mechanism. Conversely, \sage, which samples a fixed-size neighborhood, overlooks subtle structural changes, such as those resulting from elusive APT attacks.
\pname's graph structural features enable surrogate \dts to efficiently approximate \gnn models' decision-making process on in-distribution data. Although the current features are sensitive to attacker's TTPs, \pname is expandable to incorporate new features to extend support for the new TTPs.

\begin{figure*}[h]
    \centering
    \resizebox{\textwidth}{!}{
        \includegraphics[width=1.\linewidth]{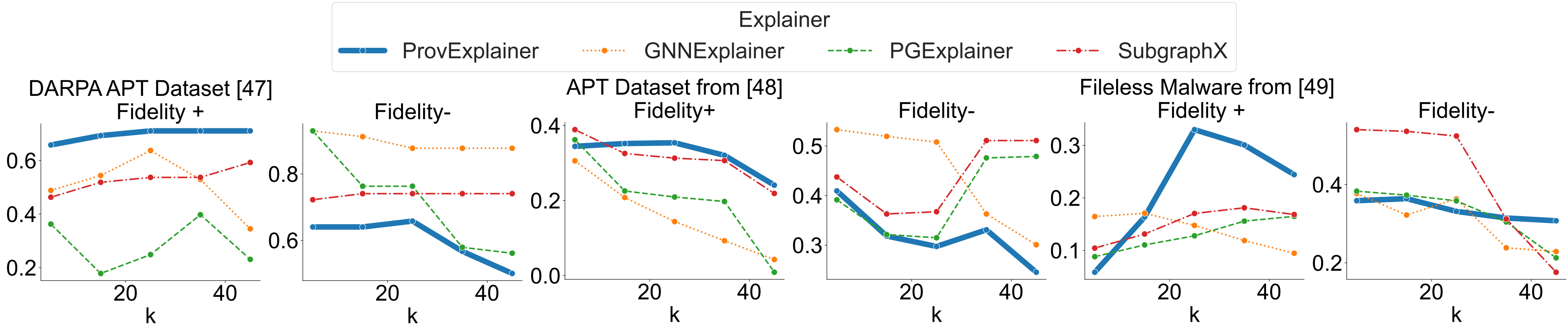}
    }
    \caption{Comparison of \gat explainers in identifying important structures using \fidelityp (higher is better) and \fidelitym (lower is better). }
    \label{fig:all_gat_fidelity}
\end{figure*}

\begin{figure*}[h]
    \centering
    \resizebox{\textwidth}{!}{
        \includegraphics[width=1.\linewidth]{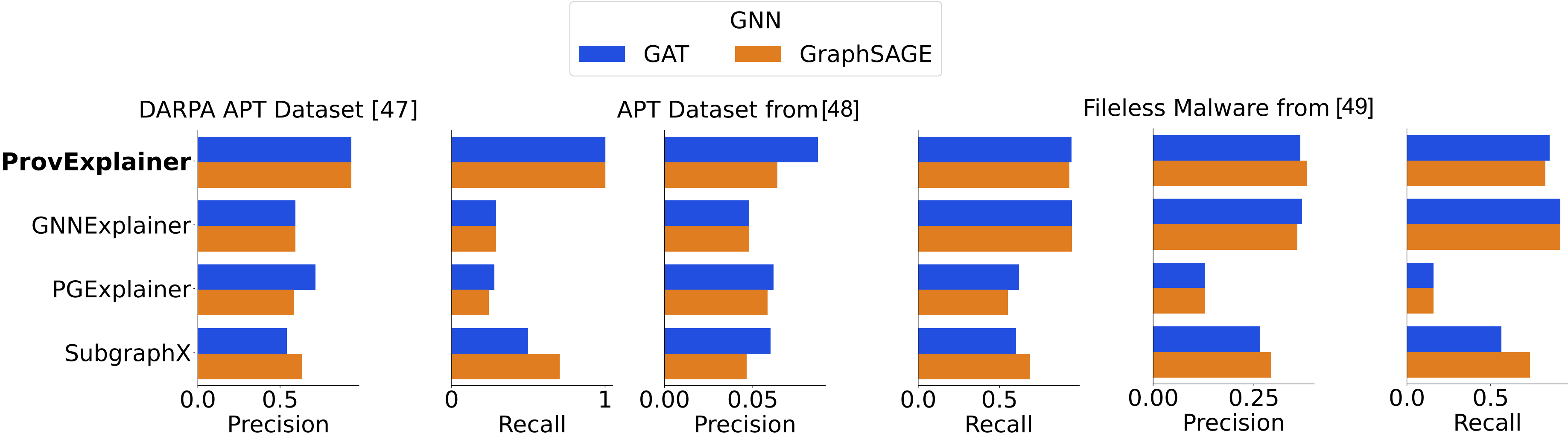}
    }
    \caption{Effectiveness of \gnn explainers at identifying documented entities, measured using precision and recall (higher is better for both).}
    \label{fig:all_pr}
\end{figure*}

\subsection{Ablation Study}\label{sec:ablation_gat}

\autoref{fig:gat_f1} shows the contributions of different features to the overall surrogate \dts agreement of approximating a \gat model. Interested readers may find this study with \sage in the appendix, \autoref{fig:all_sage_fidelity}. We separate the features into two groups: subgraphs and graph structural features. We evaluate each group individually and then we look at the their combined effect. We further separate the group into individual features (\eg top-$K$ subgraphs) and look at the effect of each constituting features in the group. Furthermore, to explore the performance of different groups, we evaluate different graph structural features individually and collectively. 

In the APT dataset, the top-$K$ subgraphs individually show different levels of agreement for both the Enterprise and Supply-Chain subsets, indicating that distinct subgraphs capture complementary signals. A similar trend is observed for the \darpatc and \fm datasets, although the increase in agreement levels plateaus as $K$ approaches 4. This suggests that the highest-ranked subgraphs encode the most informative patterns, supporting our claim that the decision path extracted from the surrogate \dt effectively captures the discriminative power of the \gnn.

We examine individual structural features, certain metrics (\ie eigenvector centrality and clustering coefficient) consistently exhibit high agreement in the \apt datasets, particularly in \darpatc (Trace and Theia) dataset. In contrast, other metrics (\ie closeness and betweenness centrality) show low agreement across most \apt datasets. This behavior arises because, in the context of \apt attacks, centrality measures often encode information that misleads the model: benign graphs contain values similar to those of attack graphs. As attackers commonly leverage system processes~\cite{darpa} to do their bidding and system processes exhibit similar structural characteristics, which reduces their discriminative power. 

In the \apt dataset, where attackers employ stealthy techniques, which suggests the necessity of integrating subgraph structures with graph structural features to construct a comprehensive feature set as combined features provide stronger alignment than relying on individual features. \pname achieves the best performance across all datasets when all features are combined.

An interesting pattern is that the \fm dataset consistently performs exceptionally well, particularly with individual features like eigenvector centrality and clustering triangles, which achieve near-perfect alignment scores. This suggests that the structural properties in the Fileless Malware dataset are distinctively different compared to the benign programs they are masquerading. 


\subsection{\pname vs. \sota Explainers} \label{sec:comp_sota}

To answer \textbf{RQ2}, \autoref{fig:all_gat_fidelity} compares the \fidelityp/\fidelitym and \autoref{fig:all_pr} compares the precision/recall of the explanations derived from \pname and \sota methods (\gnne~\cite{gnnexplainer}, \pge~\cite{pgexplainer}, and \subx~\cite{subgraphx}). We analyze specific case studies in \autoref{sec:case-studies}. 

\heading{$\text{Fidelity}^{+}$ and $\text{Fidelity}^{-}$.} In the \apt dataset from \cite{mukherjee2023sec}, \pname achieves the best performance for both \fidelityp and \fidelitym as seen in \autoref{fig:all_gat_fidelity}. \pname and \subx show competitive \fidelityp results, outperforming both \gnne and \pge. For \fidelitym, \gnne performs the worst, as it fails to identify all the critical subgraph structures. Therefore, when graph structures deemed unimportant are removed, the predictions still change, resulting in high \fidelitym value. \pname's superior performance in the DARPA and APT datasets is attributed to its security-aware features, which provide a significant advantage in identifying and extracting critical nodes from provenance graphs. These nodes, when removed, lead to prediction changes, demonstrating \pname's ability to capture essential graph components. 

Additionally, \pname achieves high coverage of important nodes, reflected in its low \fidelitym values, removing unimportant nodes does not affect predictions. In the \fm dataset, \pname demonstrates moderate performance at lower $k$ values (\eg the total number of explanation nodes considered). However, as $k$ increases, \pname becomes the best performer in \fidelityp, with \subx ranking second. These results highlight that \subx focuses on a limited number of structures and fails to provide comprehensive coverage of all the critical substructures essential for accurate predictions.

\pname's \fidelitym performance is hindered in the \fm dataset due to the unique substructures created by \fm attacks, which are not adequately captured by its security-aware features. If distinguishing subgraphs present in the test set are absent from the training set, these critical structures are missed, leading to gaps in the explanations. However, the \fidelitym performance of other explainers declines more rapidly as $k$ increases in the \fm dataset. This suggests that general-purpose \gnn explainers, which operate by calculating masks around edges and subgraphs, perform better when unique subgraph structures are present for each specific attack. It is important to note that \pname's primary objective is not to serve as a zero-day vulnerability detector but rather to produce human-interpretable explanations.

\heading{Precision and Recall.} In the APT datasets, \pname demonstrates superior precision and recall performance for both \gat and \sage networks compared to other \gnn explainers as seen in \autoref{fig:all_pr}. This highlights \pname's reliability in generating accurate explanations for the APT datasets by identifying relevant subgraph structures that include nodes mentioned in the ground truth. A similar trend is observed in \fm when explaining \gat, although \pname achieves the second-best precision and recall in this case. These results emphasize \pname's strength in explaining complex \apt datasets, where common TTPs from the MITRE framework are frequently reused (\ie similar subgraphs are formed). However, \pname struggles with the \fm dataset, where unique structures in the testing set are absent in the training set, illustrating the impact of novel TTPs on explainer performance.

\gnne performs well in the \fm dataset category due to its fidelity-driven approach, where \gnne calculates masks for substructures for every instance, enabling it to effectively identify diverse structures. Similarly, \subx shows improved precision and recall when explaining \sage compared to \gat, as \subx focuses on identifying important subgraphs based on fixed-size node neighborhoods. These neighborhoods influence the message-passing algorithm of \sage, ensuring that the explanations generated by \subx align closely with \sage's computational behavior.

\pname's performance demonstrates its strength in the APT and \fm datasets, achieving both high precision and perfect recall. This suggests a strong alignment with the inherent characteristics and structure of these datasets. However, in the more complex APT dataset~\cite{mukherjee2023sec}, all explainers, including \pname, face challenges in maintaining high precision. This limitation arises due to the dataset's complexity and the novel TTPs employed by APT attackers during certain stages. These TTPs are represented in only a subset of the graphs. Consequently, when these specific graphs appear exclusively in the testing set, \pname lacks the corresponding subgraphs for explanation, resulting in missed patterns and a subsequent decline in precision.

\begin{table}[ht]
\centering
\caption{Explanation actionability comparison between \pname and \sota \gnn explainers measured by human actionability metric. Absolute human actionability metric (top) and relative factor versus \pname (italic, bottom). Lower values reflect better actionable.}
\label{tab:human-eval}
\resizebox{\columnwidth}{!}{
\begin{tabular}{lcccc}
\toprule
\makecell[c]{Dataset} & \gnne & \pge & \subx & \textbf{\pname} \\
\midrule
\makecell[l]{DARPA APT \\Dataset \cite{darpa}} &
\makecell{3.29 \\ \emph{(1.51$\times$)}} &
\makecell{2.60 \\ \emph{(1.20$\times$)}} &
\makecell{2.36 \\ \emph{(1.08$\times$)}} &
\makecell{2.18 \\ \emph{(1.00$\times$)}} \\
\addlinespace
\makecell[l]{APT Dataset \\from \cite{mukherjee2023sec}} &
\makecell{1.97 \\ \emph{(1.59$\times$)}} &
\makecell{1.71 \\ \emph{(1.38$\times$)}} &
\makecell{1.85 \\ \emph{(1.50$\times$)}} &
\makecell{1.24 \\ \emph{(1.00$\times$)}} \\
\addlinespace
\makecell[l]{Fileless Malware \\from \cite{survivialism2021sp}} &
\makecell{2.04 \\ \emph{(1.52$\times$)}} &
\makecell{1.74 \\ \emph{(1.29$\times$)}} &
\makecell{1.77 \\ \emph{(1.32$\times$)}} &
\makecell{1.34 \\ \emph{(1.00$\times$)}} \\
\bottomrule
\end{tabular}}
\end{table}

\subsection{Human Actionability Study}\label{sec:human_study}
To answer \textbf{RQ3}, we evaluate the actionability of \pname's explanations using the metric defined in \autoref{sec:eval-metric}. Across all datasets, \pname consistently yields the lowest absolute graph traversal distance (as shown in \autoref{tab:human-eval}), representing a relative improvement of $\sim$1.4$\times$ compared to other \sota \gnn explainers. On the APT datasets, \pname achieves up to 1.5$\times$ reductions over baselines, while on the Fileless Malware dataset it attains a 1.3$\times$ reduction. The consistent improvements across diverse real-world datasets demonstrate that \pname's explanations are substantially more actionable, enabling security analysts to identify malicious edges more efficiently during investigations.

\section{Case Studies}\label{sec:case-studies}

To demonstrate \pname in a realistic setting, we analyze case studies from the DARPA~\cite{darpa} datasets. In each study, we use an explanation size of 40 nodes and refer to the \gat model. We list the most important features from the surrogate \dt and qualitatively analyze the explanations from \pname and the \sota \gnn explainers. One more case study (\autoref{sec:case3}) and detailed analyses in the appendix (\autoref{sec:deatil_attack_desc}). 

\subsection{FiveDirections: Browser Extension~\cite{darpa}}\label{sec:case2}

\begin{figure}[!htb]
  \centering
  \begin{subfigure}{0.49\columnwidth}
    \includegraphics[width=\linewidth]{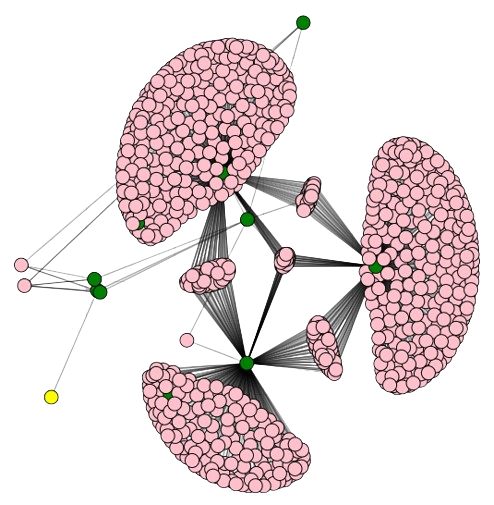}
    \caption{Attack Graph.}
    \label{fig:fived-attack}
  \end{subfigure}
  \hfill %
  \begin{subfigure}{0.49\columnwidth}
    \includegraphics[width=\linewidth]{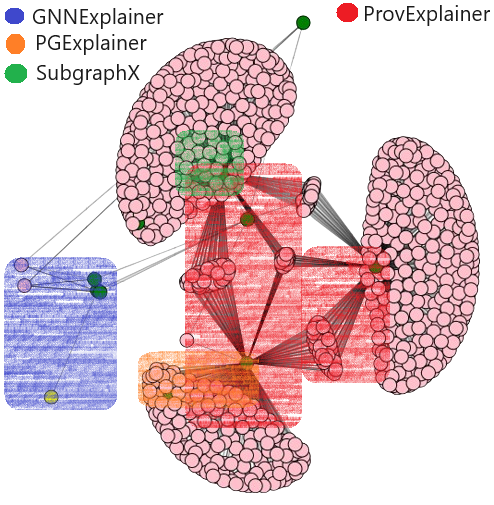}
    \caption{\sota \gnn explainations.}
    \label{fig:fived-sota}
  \end{subfigure}
  \hfill %
  \begin{subfigure}{0.70\columnwidth}
    \includegraphics[width=\linewidth]{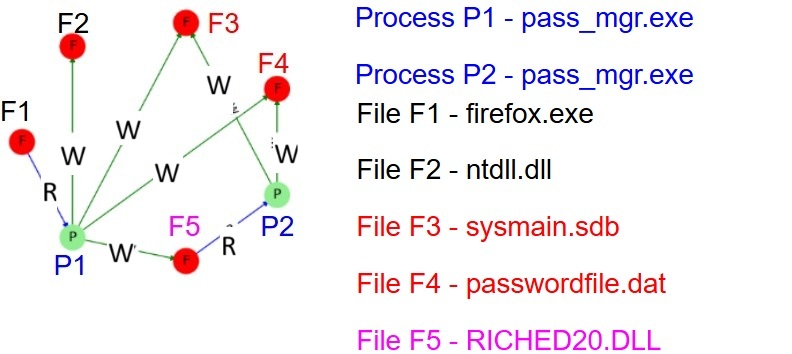}
    \caption{Pattern \#1.}
    \label{fig:fived_p1}
  \end{subfigure}
  \caption{FiveDirections: the attacker exploits the target via a malicious \code{Firefox} extension.}
  \label{fig:overall_fived}
\end{figure}

\heading{Description.}
An attacker targets the \code{Firefox} browser using a malicious extension to deploy the \code{drakon} malware. The attacker writes \code{drakon} directly to disk and exploits a compromised browser extension masquerading as a password manager to execute malicious \code{powershell} code, gaining deeper access and control over the system (\autoref{fig:overall_fived}).

\heading{Features:} Degree centrality and pattern \#1.

\heading{System Interpretation.}
The pattern shown in \autoref{fig:fived_p1} provides a detailed representation of the initial stages (\ie initial compromise and privilege escalation) of \apt attack lifecycle orchestrated by the Drakon APT, showcasing a sequence of file and process interactions. Process P1 (\code{pass\_mgr.exe}) serves as the initial pivot point, leveraging File F1 (firefox.exe) to gain access to the system through a compromised browser backdoor. This initial stage involves P1 writing to multiple files, including File F2 (\code{ntdll.dll}), File F3 (\code{sysmain.sdb}), File F4 (\code{passwordfile.dat}), and File F5 (\code{RICHED20.DLL}). These operations signify the deployment of malicious shellcode, preparation for privilege escalation, and establishment of a foothold within the system.

The transition to Stage 2 is marked by the activities of Process P2 (\code{pass\_mgr.exe}), which reads from File F5 (\code{RICHED20.DLL}) and writes to critical files such as File F3 and File F4. These interactions indicate the progression to privilege escalation and credential harvesting, leveraging tools like Copykatz and Drakon APT's elevate driver. The repeated use of the same process executable (\code{pass\_mgr.exe}) across different stages points to sophisticated \apt \ttps such as process injection or executable reuse.

Since the attack graph contains high quantities of file read and write flowers for a relatively small number of nodes between these flowers, it leads to a high degree of centrality, which means a high amount of process interaction between static artifacts like files. The graph's structural measures complement the subgraph's interpretation and identifies the TTPs employed by Drakon APT.

\heading{\pname vs. \sota explainers.}
\autoref{fig:fived-sota} compares the explanations of \pname and those of \sota \gnn explainers, focusing on their efficacy in identifying security-aware elements. \pname excels by highlighting the malware replicating itself from template files and accessing sensitive system files. \pname isolates security-relevant structures in the graph, significantly increasing the end-user trust in the detection.

\gnne identifies the malware template file and the malicious extension. This effectiveness stems from \gnne's method of searching for important edges. When this \gnne isolates the pivot structure, the graph becomes disjoint, leading to a change in prediction. This results in a high information gain, the core metric for \gnne. On the other hand, \pge and \subx reveal commonalities across attack graphs, such as the identification of key system library accesses required for malware operation. \subx's effectiveness varies due to its Monte Carlo Tree Search (MCTS), suggesting benefits to incorporating domain-specific insights into \subx's scoring function.

\section{Discussion and Future Work}\label{sec:discussion-and-conclusions}

\heading{Limitations.} 
\pname depends on previously observed graph patterns, as the surrogate DT is trained on discriminant subgraph patterns mined from an existing dataset. Therefore, \pname is unlikely to perform well on attacks that contain previously unseen subgraph patterns. We recognize the inherent limitations of our approach and, therefore, set explicit boundaries for \pname's interpretive range, aiming to validate detection results and provide actionable intelligence for known attacks. Preconditions that may prevent ProvExplainer from explaining some attacks: \nm{1} the detection model must detect the attack, \nm{2} the attack graph contains distinct subgraphs identified during the mining phase, and \nm{3} the attack's distinctive subgraphs are small enough to be feasibly pattern-matched.

\heading{Adversarial Manipulation of Graph Features.}
The explainer models used in \pname is separate from the black-box detection model. Fooling both the surrogates and the detection model is the ideal case for an attacker seeking to avoid suspicion, and an attacker who can attack the explainer this way can also attack the detection model directly~\cite{goyal2023sometimes, mukherjee2023sec}. Creating robust explanation framework that can withstand adversarial manipulation is a critical open research problem, and is beyond the scope of this work.

\section{Conclusion}\label{sec:conc}
We introduced \pname, a framework designed to address a critical gap in explainable ML for cybersecurity. It provides transparent, interpretable insights into \gnn-based \ids in the system provenance domain. By leveraging security-aware discriminative subgraph patterns and graph structural features, \pname facilitates instance-level explanations that consider both local and global graph properties. This dual-level interpretability is essential for removing the opacity and help with the adoption of \gnn-based \ids.
\pname is a critical advancement in explainable AI for the security domain and a scalable framework for real-world deployment.

\printbibliography{}

\appendix
\section{Appendix}\label{sec:appendix}

\begin{table}
    \caption{Node and edge types for in-house system provenance with associated attributes.}
    \label{tab:schema}
    \centering
    \resizebox{.9\linewidth}{!}{
        \begin{tabular}{@{}lll@{}}
            \toprule
            & Types & Attributes \\ 
            \midrule
            
            \multirow{3}{*}{\makecell{Nodes\\(resources)}} 
                & process & signature, executable name, pid \\
            & file & owner (uid, gid), name, inode        \\
            & socket (IP)   & dstip, srcip, dstport, srcport, type \\ 
            \midrule
            \multirow{3}{*}{\makecell{Edges\\(events)}} 
                & process $\rightarrow$ process    & command args, starttime \\
            & process $\rightarrow$  file       & read, write, amount \\
            & process $\rightarrow$ IP address & send, recv, amount \\ 
            \bottomrule
        \end{tabular}
    }
\end{table}

\subsection{Implementation}\label{sec:data_collection}

Our data collection module can operate on Windows systems by using the Windows ETW~\cite{winETW} and on Linux system by using Linux auditd~\cite{audit} frameworks to collect relevant system calls regarding files, processes and network sockets. These include system calls for \nm{1} file operations (\eg {\tt read()}, {\tt write()}, {\tt unlink()}), \nm{3} network socket operations (\eg {\tt connect()}, {\tt accept()}), \nm{4} process operations (\eg {\tt create()}, {\tt exec()}, and {\tt exit()}).
The system-level data is stored in a PostgresSQL database. 
\autoref{tab:schema} summarizes the event collection schema~\wc{specification} for our in-house deployment. While it includes major nodes and event types annotated with textual and numerical attributes, detailed entries of the event collection differ across OS environment. 

\pname is implemented using \code{python} and graph generation framework is implemented in \code{java}. The \gnn training and evaluation pipelines are implemented using Deep Graph Library (DGL)~\cite{dgl} framework, and the surrogate \dts are implemented using sklearn~\cite{scikit}. We used the frequent subgraph miner, GERM~\cite{berlingerio2009mining}, to mine patterns with a minimum support of 1 and a maximum subgraph size of 10. The discriminant subgraph patterns were selected by maximizing the scoring function. The scoring function's hyperparameters and DT were empirically tuned using grid search by iteratively evaluating \pname's performance using Fidelity+/- and Precision/Recall metric on cross-validation folds.

\subsection{Dataset Statistics\wc{Demographics}}\label{sec:dataset_stats}


\newcommand{\ExplorerAnomalousNodes}{145.4}
\newcommand{\ExplorerAnomalousEdges}{166.3}
\newcommand{\ExplorerAnomalousGraphs}{673}

\newcommand{\ExplorerBenignNodes}{37.1}
\newcommand{\ExplorerBenignEdges}{37.7}
\newcommand{\ExplorerBenignGraphs}{1057}

\newcommand{\NetshBenignNodes}{34.2}
\newcommand{\NetshBenignEdges}{44.1}
\newcommand{\NetshBenignGraphs}{1023}

\newcommand{\NetshAnomalousNodes}{117.9}
\newcommand{\NetshAnomalousEdges}{122.5}
\newcommand{\NetshAnomalousGraphs}{537}

\newcommand{\NsLookupBenignNodes}{42.8}
\newcommand{\NsLookupBenignEdges}{50.3}
\newcommand{\NsLookupBenignGraphs}{879}

\newcommand{\NsLookupAnomalousNodes}{100.8}
\newcommand{\NsLookupAnomalousEdges}{126.1}
\newcommand{\NsLookupAnomalousGraphs}{281}

\newcommand{\PythonBenignNodes}{26.2}
\newcommand{\PythonBenignEdges}{32.3}
\newcommand{\PythonBenignGraphs}{3621}

\newcommand{\PythonAnomalousNodes}{168.4}
\newcommand{\PythonAnomalousEdges}{227.4}
\newcommand{\PythonAnomalousGraphs}{822}

\newcommand{\RundllBenignNodes}{39.6}
\newcommand{\RundllBenignEdges}{46.5}
\newcommand{\RundllBenignGraphs}{2462}

\newcommand{\RundllAnomalousNodes}{145.5}
\newcommand{\RundllAnomalousEdges}{144.9}
\newcommand{\RundllAnomalousGraphs}{821}

\newcommand{\SchtasksBenignNodes}{14.3}
\newcommand{\SchtasksBenignEdges}{19.7}
\newcommand{\SchtasksBenignGraphs}{1429}

\newcommand{\SchtasksAnomalousNodes}{134.9}
\newcommand{\SchtasksAnomalousEdges}{223.0}
\newcommand{\SchtasksAnomalousGraphs}{329}

\newcommand{\SvchostBenignNodes}{5.62}
\newcommand{\SvchostBenignEdges}{4.76}
\newcommand{\SvchostBenignGraphs}{1147}

\newcommand{\SvchostAnomalousNodes}{176.4}
\newcommand{\SvchostAnomalousEdges}{220.6}
\newcommand{\SvchostAnomalousGraphs}{734}

\begin{table}[!tb]
    \caption{\apt and Fileless Malware graph statistics.}
    \label{table:anomaly_vertex_edge}
    \center
    \resizebox{1.0\columnwidth}{!}{%
    \begin{tabular}{@{}lcccc@{}}\toprule
        
        Applications & \makecell[c]{\# of Benign \\ Graphs} & \makecell[c]{\# of Anomaly \\ Graphs} & \makecell[c]{Avg \# of Benign\\ Nodes / Edges} & \makecell[c]{Avg \# of Anomaly\\ Nodes / Edges}\\
        \midrule
        \multicolumn{5}{c}{DARPA APT Dataset~\cite{darpa}} \\
        \midrule
        \code{Trace} & 1883 & 8 & 735.35 / 957.56 & 836.15 / 946.75\\
        \code{Theia} & 2858 & 9 & 559.47 / 979.59 & 913.91 / 987.31\\
        \code{FiveDirections} & 954 & 13 & 906.22 / 971.91 & 959.43 / 973.63\\
        \midrule
        \textbf{Average} & \textbf{1898.33} & \textbf{10.00} & \textbf{669.81 / 982.18} & \textbf{777.08 / 1011.19} \\
        \midrule
        \multicolumn{5}{c}{APT Dataset from \cite{mukherjee2023sec}} \\
        \midrule
        \code{Enterprise} & 3079 & 1836 & 90.22 / 85.13 & 73.73 / 76.88 \\
        \code{Supply-Chain} & 3212 & 1092 & 65.02 / 40.77 & 61.09 / 54.33\\
        \midrule
        \textbf{Average} & \textbf{3145.50} & \textbf{1464.00} & \textbf{55.94 / 83.29} & \textbf{58.38 / 51.69} \\
        \midrule
        \multicolumn{5}{c}{Fileless Malware from \cite{survivialism2021sp}} \\
        \midrule
        \code{wscript.exe} & 399 & 40 & 66.18 / 59.94 & 44.10 / 56.45 \\
        \code{cscript.exe} & 876 & 11 & 88.56 / 81.58 & 87.36 / 101.54 \\
        \code{reg.exe} & 309 & 116 & 60.18 / 52.93 & 78.91 / 131.87 \\
        \code{python.exe} & 15585 & 426 & 89.95 / 83.15 & 57.98 / 79.33 \\
        \code{explorer.exe} & 399 & 40 & 66.18 / 59.94 & 44.10 / 56.45 \\
        \code{netsh.exe} & 621 & 7 & 44.08 / 37.49 & 38.42 / 52.28 \\
        \code{net.exe} & 621 & 7 & 44.08 / 37.49 & 38.42 / 52.28 \\
        \code{rundll32.exe} & 1632 & 443 & 52.42 / 46.42 & 51.97 / 81.31 \\
        \code{mshta.exe} & 876 & 11 & 88.56 / 81.58 & 87.36 / 101.54 \\
        \code{schtasks.exe} & 399 & 40 & 66.18 / 59.94 & 44.10 / 56.45 \\
        \code{conhost.exe} & 309 & 116 & 60.18 / 52.93 & 78.91 / 131.87 \\
        \code{svchost.exe} & 1632 & 443 & 52.42 / 46.42 & 51.97 / 81.31 \\
        \midrule
        \textbf{Average} & \textbf{2667.00} & \textbf{90.84} & \textbf{60.44 / 54.15} & \textbf{63.72 / 89.93} \\
        \bottomrule
    
    \end{tabular}
    }
       
\end{table}

%
%
%

\heading{Anomaly Detection Dataset.} 
The dataset statistics for the anomaly detection dataset is seen in \autoref{table:anomaly_vertex_edge}. The Fileless Malware samples were downloaded from \cite{virustotal} and selected from recent studies, \cite{grammatikakis2021csr,kuchler2021does,avllazagaj2021malware,survivialism2021sp}. The benign provenance graphs for the anomaly detection dataset were sourced from the DARPA dataset, which contained 5,695 benign graphs with an average of 669.81 vertices and 982.18 edges (\autoref{table:anomaly_vertex_edge}); the \fm dataset, which contained 19,422 benign graphs with an average of 60.44 vertices and 54.15 edges; and the APT dataset, which contained 6,291 benign graphs with an average of 55.94 vertices and 83.29 edges. The corresponding anomaly graphs are sourced from the same datasets: The DARPA dataset contained 30 anomalous graphs with an average of 777.08 vertices and 1011.19 edges (\autoref{table:anomaly_vertex_edge}); the \fm contained 1,043 anomalous graphs with an average of 63.72 vertices and 89.93 edges; and the APT dataset contained 2,928 graphs with an average of 58.38 vertices and 51.69 edges.

\subsection{Ablation Study (using GraphSAGE)}
\begin{figure*}[tbh]
  \centering
  \resizebox{\textwidth}{!}{%
    \begin{minipage}{\textwidth}
      \centering
      \begin{minipage}[t]{0.49\textwidth}
        \vspace{0pt}
        \centering
        \begin{subfigure}[t]{0.99\linewidth}
          \centering
          \includegraphics[page=1,width=\linewidth]{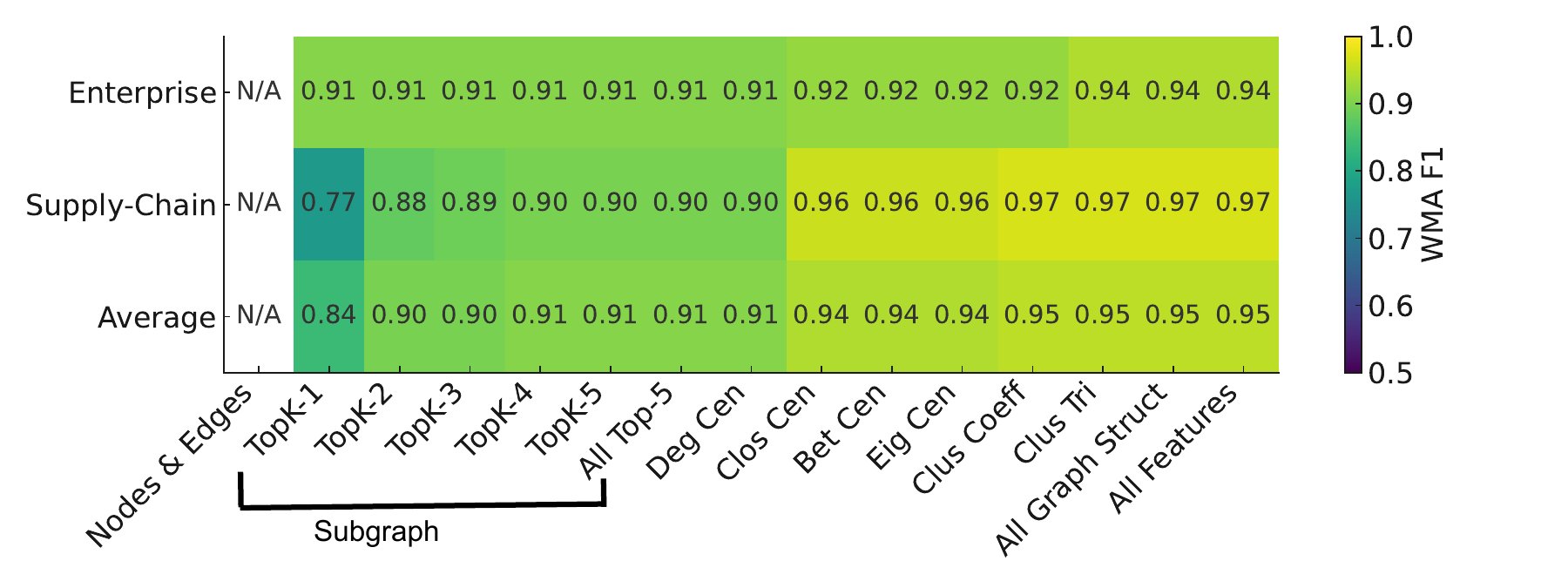}
          \caption{DARPA APT Dataset~\cite{darpa}}
          \label{fig:sage_wma_f1_heatmap_mukh}
        \end{subfigure}

        \begin{subfigure}[t]{0.99\linewidth}
          \centering
          \includegraphics[page=2,width=\linewidth]{figs/sage_wma_f1_graphs.pdf}
          \caption{APT Dataset from \cite{mukherjee2023sec}}
          \label{fig:sage_wma_f1_heatmap_darpa}
        \end{subfigure}
      \end{minipage}
      \hfill
      \begin{minipage}[t]{0.49\textwidth}
        \vspace{0pt}
        \centering
        \begin{subfigure}[t]{0.99\linewidth}
          \centering
          \includegraphics[page=3,width=\linewidth]{figs/sage_wma_f1_graphs.pdf}
          \caption{Fileless Malware from \cite{survivialism2021sp}}
          \label{fig:sage_wma_f1_heatmap_fileless}
        \end{subfigure}
      \end{minipage}
    \end{minipage}%
  }
  \caption{Agreement (higher is better) measured using WMA F1 of surrogate decision trees versus the \sage model across different feature sets.}
  \label{fig:sage_f1}
\end{figure*}

The ablation study utilizing the \sage network revealed patterns consistent (as shown in ~\autoref{fig:sage_f1}) with those observed in the ablation study using \gat network~\autoref{sec:ablation_gat}. It was again noted that for \fm datasets the graph structural features provided better feature selection than subgraph patterns. This pattern is observed because the malware graphs are distinctively different when compared against their benign counter part. Finally, the pattern that subgraph patterns with the graph structures provided the best feature selection.

\subsection{\pname vs. \sota Explainers}

\begin{figure*}[!htb]
    \centering
    \resizebox{\textwidth}{!}{
        \includegraphics[width=1.\linewidth]{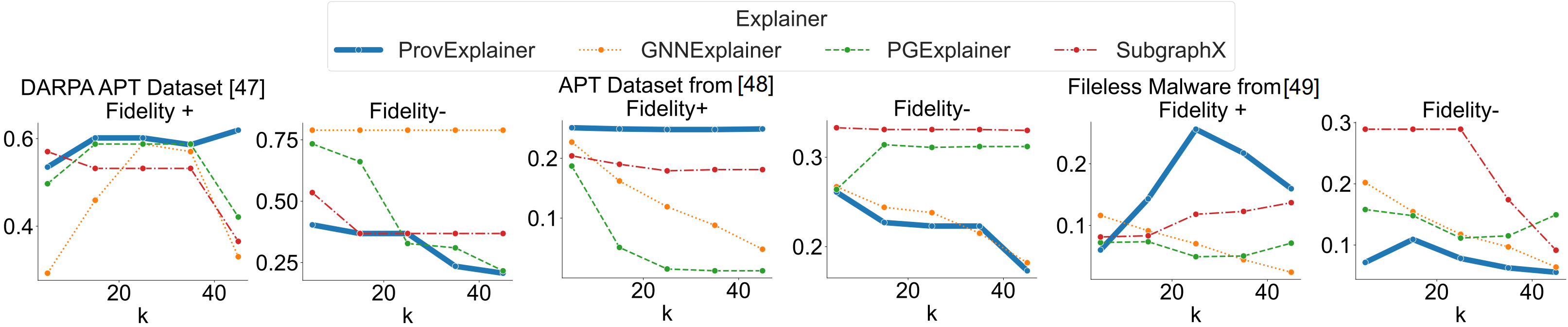}
    }
    \caption{Effectiveness of \sage explainers in identifying prediction-relevant graph structures using \fidelityp and \fidelitym.}
    \label{fig:all_sage_fidelity}
\end{figure*}

Similar to the \gat-based results, \pname demonstrates superior performance across various datasets. The complexity of the provenance graphs, defined by their size and the nature of resource interactions, significantly influences the effectiveness of different explainers on specific datasets. For instance, on provenance graphs such as the \apt graphs from \cite{mukherjee2023sec}, \pge achieves the second-best performance. However, for the more complex or noisy DARPA \apt graphs, \gnne emerges as the second most effective method. 


\subsection{Case Study: FiveDirections (Copykatz)~\cite{darpa}}\label{sec:case3}

\begin{figure}[!htb]
  \centering
  \begin{subfigure}{0.49\columnwidth}
    \includegraphics[width=\linewidth]{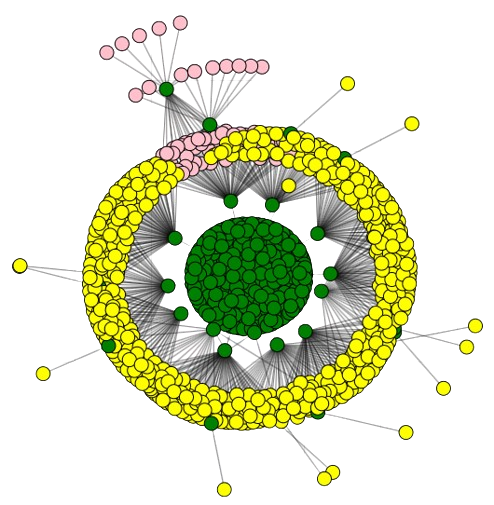}
    \caption{Attack Graph.}
    \label{fig:fived2-attack}
  \end{subfigure}
  \hfill %
  \begin{subfigure}{0.49\columnwidth}
    \includegraphics[width=\linewidth]{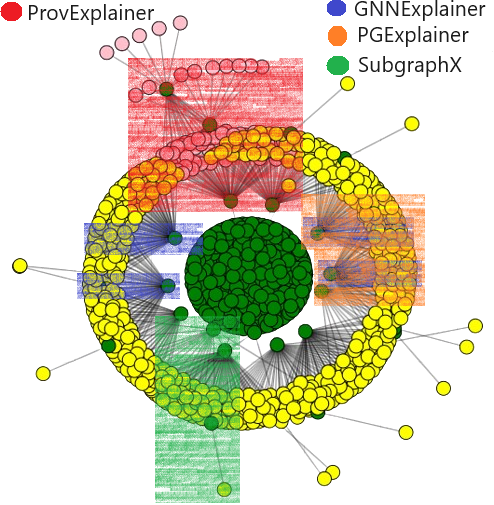}
    \caption{\sota \gnn explainations.}
    \label{fig:fived2-sota}
  \end{subfigure}
  \hfill %
  \begin{subfigure}{0.90\columnwidth}
    \includegraphics[width=\linewidth]{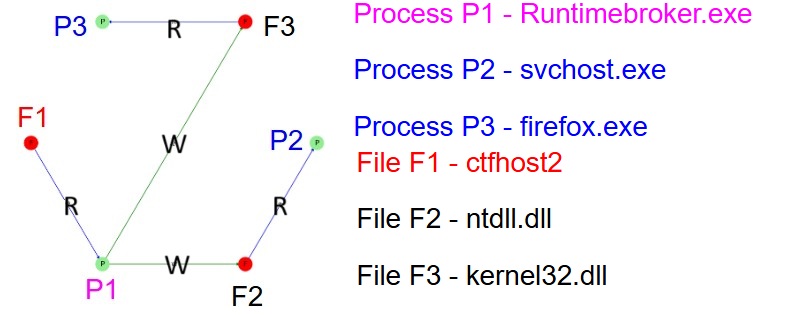}
    \caption{Pattern \#2.}
    \label{fig:fived2_p2}
  \end{subfigure}
  
  \caption{FiveDirections: the attacker gains C2 connections and installs \code{Copykatz} through a \code{Firefox} exploit.}
  \label{fig:overall_fived2}
\end{figure}

\heading{Description.}
In a sophisticated attack, a hijacked version of \url{usdoj.gov} exploits \code{Firefox} to deploy \code{drakon} malware to the victim host. Then, \code{drakon} uses the \textit{elevate} driver to escalate privileges and masquerade as the \code{runtimebroker.exe} system program. Finally, the malicious \code{runtimebroker.exe} instance connects to a command and control (C2) server to download and execute \code{Copykatz} (an older version of \code{Mimikatz}) to harvest and exfiltrate host credentials, as illustrated in \autoref{fig:overall_fived2}.

\heading{Features:} Degree centrality and pattern \#2.

\heading{System Interpretation.}
The pattern shown in \autoref{fig:fived2_p2} highlights a multi-stage attack sequence characteristic of the Firefox Drakon APT focusing on elevating the privilege using \code{Copykatz}. Process P3 (\code{firefox.exe}) initiates the attack by exploiting its backdoor to download the payload, File F1 (\code{ctfhost2.exe}) that contains the \code{drakon} malware. This payload is subsequently executed by Process P1 (\code{runtimebroker.exe}), which modifies critical system libraries, File F2 (\code{ntdll.dll}) and File F3 (\code{kernel32.dll}). These modifications indicate the preparation of the system environment for privilege escalation and sustained malicious operations. Process P2 (\code{svchost.exe}) interaction with File F2 emphasizes its role in leveraging system-level components to propagate the attack, consistent with the exploitation of driver signing for privilege elevation.

The graph structure captures the interdependencies between processes and system-critical files, illustrating the deliberate flow of the attack. By chaining interactions, such as \code{runtimebroker.exe} modifying \code{ntdll.dll} and \code{svchost.exe} subsequently accessing it, the graph provides insights into how the malware manipulates system operations. Furthermore, the reuse of Firefox as both an initial access vector and a conduit for \code{kernel32.dll} interactions highlights the advanced tactics employed to achieve persistence and exploitation. These observations align closely with the described TTPs of the Firefox Drakon Copykatz APT, offering a clear mapping of graph elements to specific stages of the attack lifecycle. The attack graph contains high quantities of file read and writes flowers for a relatively small number of process nodes acting as bridges between the flowers, so it leads to a high degree of centrality.

\heading{\pname vs. \sota explainers.}
\gnne and \pge pinpoint the stage where multiple \code{Firefox} processes connect to the C2 servers. This activity traces back to the malicious \code{runtimebroker} instance. Notably, \subx detects an alternate trajectory where a \code{Firefox} process, instead of reaching out to external servers, spawns another process aimed at local content manipulation, showcasing the malware's versatility in engaging with both external and internal resources for its objectives. While \sota explainers have demonstrated proficiency in identifying the final stages of this data breach, only \pname effectively captured both the initial infection and its privilege escalation attempts. This distinction highlights the importance of recognizing early-stage indicators for root cause analysis.

\subsection{In-Depth Analysis of Case Studies}\label{sec:deatil_attack_desc}

\heading{Trace: Phishing E-mail~\cite{darpa}.}
The pattern \#2 captured the staging behavior of the Trojan. The Trojan was downloaded from \url{www.nasa.ng}, executed and replicated itself within the system. Specifically, the malware named \code{nasa.ng} is placed in \path{/home/admin/.mozilla/firefox/} and \path{/usr/local/firefox-54.0.1/obj-x86_64-pc-linux-gnu/}. The Trojan created new malware with benign names such as \code{firefox} to effectively evade detectors, to replicate unhindered and overloaded the system with malicious processes. After the malware successfully staged, the malware masquerading as \path{/bin/sh} to read the malicious script staged in (\path{/etc/update-motd.d/00-header/}) and executed it to create multiple copies of itself. The malware reads various system configuration files present in \path{/etc/protocols/}, \path{/etc/lsb-release/}, and \path{/etc/hosts.deny/}. Reading sensitive system configuration files are essential to build the system profile. The pattern \#5 captured the dependency correlation of the malware  created that inherited configurations.

Ultimately, the malware completes its target of reading sensitive system files from \path{/etc/fonts/conf.d/}, \path{/usr/lib/x86_64-linux-gnu/}, and \path{/usr/share/X11/local/}. These activities are aimed at gathering system information to create a profile of the company and the devices in use. The attacker wants to create a profile of the victim environment to ensure their malware can effectively leverage system libraries to complete their objective. There is an overlap in the files (present in \path{\etc\hosts} and \path{/usr/lib/x86_64-linux-gnu/*}) involved in the pattern \#2 and pattern \#5 operations because the malware replicates itself probing and inheriting the functional dependencies of its parent. 

\gnne was able to correctly capture the staging behavior where the malware from \code{nasa.ng} read shared library (\path{/usr/lib/x86_64-*/*.so.*}) and cache file (\path{/usr/share/applications/mimeinfo.cache}, \path{/usr/lib/x86_64-*/*/loaders.cache}). \pge incorrectly indicated benign substructures, but \subx correctly captured the inheritance behavior of \code{firefox.exe} executing multiple times with the argument file \url{http://www.nasa.ng/}, to create the malware clones from the template. \sota explainers missed the malware's ultimate goal of reading sensitive files, which was only captured by \pname.

\heading{FiveDirections: Browser Extension~\cite{darpa}.}
In the context of the described attack, where the \code{drakon} malware exploits the \code{firefox.exe} browser through a rogue extension (\code{pass\_mgr}) the pattern \#1 is created. Multiple instances of the malware process are created, each reading from the malicious files: \code{passwordfile.dat} and \code{pass\_mgr.exe}. Additionally, they access essential dictionary files (\code{en-US.aff}) and cryptographic libraries (\code{bcryptprimitives.dll} ) to maintain operational consistency, allowing them to conduct their malicious activities efficiently. The  pattern \#2 captures the data extraction behavior through sensitive file accesses. These malware access sensitive information and system configuration files like \code{WindowsShell.Manifest}, and \code{wintrust.dll}, along with initial malware files (\code{addons}, \code{tzres.dll}, \code{userenv.dll}).

\gnne effectively identifies the malware template file present in \path{C:\Program Files\Mozilla Firefox\add-on\pass_mgr.exe} and the initial malware \code{pass\_mgr.exe}. \pge recognized the structures common across all attack graphs, particularly identifying file access of system libraries needed for malware operation \path{C:\*\System32\driver}, \path{C:\*\Windows\SysWOW64} and \path{C:\*\AppData\Local\Temp}. \subx performed at the same capacity as \pge as it identified a different the structure of malware executing its payload from \path{C:\*\Desktop\*\add-on}, reading sensitive files from \path{C:\*\Program Files\Mozilla Firefox\*}, and extracting them through \path{C:\*\admin\AppData\*}. This is partly due to its foundation on Monte Carlo Tree Search (MCTS), incorporating nondeterministic exploitation and exploration stages. In the absence of attribute information in the \subx, the exploitation stage lacks guidance. 

\heading{FiveDirections: Copykatz~\cite{darpa}.}
The pattern \#2 starts its kill chain stage of initial access~\cite{apts1} by writing and executes a malware masquerading as \code{firefox.exe} which contains the \code{Mimikatz} and \code{Copykatz} modules. \texttt{ntdll.dll}~\cite{ntdll} includes multiple kernel-mode functions which enables the ``Windows Application Programming Interface (API)'' and \texttt{bcryptprimitives.dll}~\cite{bcryptprimitives} contain functions implementing cryptographic primitives, which are essential for \code{Mimikatz}~\cite{mimikatz} and \code{Copykatz}.  

The \code{firefox.exe} malware write its payload into temporary files, such as \texttt{ctfhost.exe}, \texttt{virtuous} and \texttt{tropical}. These files are then executed by malware in the system creating pattern \#5. Subsequently, another malicious instance of the \code{firefox.exe} malware initiates a chain or read-execute cycle by executing the malware template as well as making external IP connections. This dependency correlation between the malware and its parent is characterized by inheritance, where the children require the same system library files as the parent to function correctly. The DLLs involved are read from \path{Program Files} and \path{System32}.

\gnne and \pge correctly identified the data exfiltration stage where four malicious \code{firefox.exe} make C2 connections to the external IPs (\url{202.179.137.58} and \url{217.160.205.44}). \code{firefox.exe} was invoked by malware masquerading as a popular system program, \code{runtimebroker.exe}. \gnne correctly masks out the process creation edges of the children \code{firefox.exe}, to create disjoint graphs and leading to prediction change. \subx also highlighted \code{firefox.exe} processes that are created by \code{runtimebroker.exe}, but instead of making external connection it creates other \code{firefox.exe} for rendering content from localhost (\url{127.0.0.1}).
\balance

\end{document}